\documentclass[10pt,aps,prl,twocolumn,superscriptaddress,nofootinbib,floatfix]{revtex4-2}
\usepackage{graphicx}
\usepackage{amsmath,amssymb}
\usepackage{natbib}
\usepackage{array}[=2016-10-06]
\usepackage{xcolor}
\usepackage{hyperref}
\begin{document}

\title{Same Events, Weaker Bound:\\ Model Dependence in Magellanic Microlensing Limits on Primordial Black Holes}

\author{David Morales-Zapien}
\affiliation{Department of Physics \& Astronomy, California State University Long Beach, Long Beach, CA 90840, U.S.A.}

\author{Lucas Brown}
\affiliation{Santa Cruz Institute for Particle Physics, 1156 High St., Santa Cruz, CA 95064, USA}
\affiliation{Department of Physics, University of California Santa Cruz, Santa Cruz, CA 95064, USA}

\author{Stefano Profumo}
\affiliation{Santa Cruz Institute for Particle Physics, 1156 High St., Santa Cruz, CA 95064, USA}
\affiliation{Department of Physics, University of California Santa Cruz, Santa Cruz, CA 95064, USA}
\date{\today}

\begin{abstract}
OGLE microlensing of the Magellanic Clouds sets the leading bounds on primordial black holes (PBHs) as dark matter from $10^{-8}$ to $10^{3}\,M_\odot$. Holding the detected events and survey efficiencies fixed, we show that changes to the Milky Way halo, the stellar lens populations and the Large Magellanic Cloud halo, each motivated by recent data, weaken the published strict 20-year bound by up to $\simeq6$, and by over an order of magnitude in an extreme stress test. With primordial clustering, the planetary-mass bound weakens by a factor $3.4$--$6$. We strongly recommend that compact-object dark matter limits be quoted together with their modeling assumptions and, generally, be derived under conservative assumptions.
\end{abstract}

\maketitle

\paragraph{Introduction.}
Primordial black holes (PBHs) \citep{Zeldovich1967,Hawking1971,CarrHawking1974} remain a viable dark-matter (DM) candidate over a wide range of masses \citep{CarrKuhnel2020,Carr2021RPP,GreenKavanagh2021}. Interest in two mass windows has been renewed by gravitational-wave detections of black-hole mergers and by the excess of short-timescale microlensing events toward the Galactic bulge \citep{Niikura2019Bulge}. In both windows the decisive test is gravitational microlensing \citep{Paczynski1986} of the Magellanic Clouds. The OGLE surveys \citep{Wyrzykowski2011LMC,Wyrzykowski2011SMC,Mroz2024Nature,Mroz2024ApJL,Mroz2025ApJS}, following EROS-2 \citep{Tisserand2007}, limit PBHs to at most a few percent of the DM over most of the range $10^{-8}$--$10^{3}\,M_\odot$, and these curves are routinely used as fixed inputs in cosmological and particle-physics analyses. Their robustness has recently been disputed \citep{Hawkins2025} and defended \citep{Mroz2025Reply}; the defense argues that plausible changes of the Milky Way (MW) halo profile alter the halo optical depth toward the Large Magellanic Cloud (LMC) by only a few percent.

A microlensing exclusion depends on two distinct inputs: the predicted event rate at fixed abundance $f_{\rm PBH}$, and the treatment of the detected events. OGLE already brackets the second, publishing \emph{strict} limits, which assume that no detected event is a PBH, and \emph{relaxed} limits, in which stellar and PBH lenses both contribute to the detections; the two differ by at most a factor $\simeq4$, near $0.25\,M_\odot$ \citep{Mroz2024Nature}. Previous studies have propagated halo-model uncertainties into microlensing event rates \citep{Calcino2018} and compared halo optical depths \citep{Mroz2025Reply}. Here we extend the comparison to the OGLE field-level data and timescale likelihood, varying halo and stellar-population ingredients together. Because our likelihood treats event origin as OGLE's relaxed analysis does, ratios to the relaxed curves isolate the rate model.

We find that (i) at fixed data the 20-year bound weakens, relative to OGLE's relaxed limit, by up to $\simeq1.6$ for a benchmark built from recent Galactic and LMC measurements, and by up to $\simeq3.4$ for an extreme stress test (by up to $6.2$ and $13.2$ relative to the strict limit, a comparison that also includes the event-origin difference); (ii) the high-cadence planetary-mass bound weakens by $1.2$--$2.4$; (iii) although LMC-induced disequilibrium is invoked qualitatively as a reason to distrust equilibrium halo models \citep{Hawkins2025}, it raises, rather than lowers, the lensing-weighted MW-halo density; and (iv) if PBHs are primordially clustered, the planetary-mass bound at its tabulated minimum weakens by $\simeq3.4$ (benchmark) to $\simeq6.0$ (stress test) relative to the published limit once a Ly$\alpha$ bound on cluster shot noise is imposed, whereas the 20-year bound changes by at most $16\%$. Our parameters are fixed rather than marginalized, so the benchmark quantifies sensitivity; it is not a statistical lower envelope of allowed event rates.

\paragraph{Data.}
We use the 20-year OGLE-III$+$IV LMC survey \citep{Mroz2024Nature}: per-field source counts, baselines of $4000$ or $7000$~d, published detection efficiencies, and the timescale posteriors of the 13 pipeline-selected events \citep{Mroz2024ApJS}. We also reanalyze the high-cadence campaign \citep{Mroz2024ApJL}: five LMC and four Small Magellanic Cloud (SMC) fields with $\simeq35$ million sources, one long-timescale event, and efficiencies $\epsilon_j(t_E,\rho)$ injected over a $700$~d window. As in the published analysis, we sum the expected events from both clouds and neglect the SMC halo, so that the comparison is like for like (see \ref{sec:hc_scope} of the Supplemental Material, SM). We do not reanalyze the long-timescale SMC sample, which does not affect the most stringent limits \citep{Mroz2025ApJS}.

\paragraph{Rate model.}
The published LMC analyses adopt a contracted, rotation-curve-calibrated MW halo \citep{Cautun2020} and an untruncated Hernquist LMC halo \citep{Mroz2024Nature,Mroz2024ApJL,Erkal2019}. Table~\ref{tab:assumptions} (SM) lists every change we make.

\emph{Milky Way halo.} We adopt the flattened double power law of \citet{Ibata2024},
\begin{equation}
\rho(s)=\rho_0\left(\frac{s}{r_0}\right)^{-\gamma}\left(1+\frac{s}{r_0}\right)^{\gamma-\beta}e^{-s^2/r_t^2},
\label{eq:mwhalo}
\end{equation}
with $s^2=R^2+z^2/q_h^2$, $\gamma=0.97$, $\beta=3$, $r_0=14.7$~kpc and $r_t=1000$~kpc. At fixed local density, smaller $q_h$ lowers the density off the Galactic plane, through which the LMC sightline passes. Gaia-era analyses favor $q_h\simeq0.75$--$0.78$ \citep{Ibata2024,Bienayme2024}, and a stream-based fit prefers a comparably flattened but tilted halo \citep{NibauerBonaca2025}. Our \emph{benchmark} takes $\rho_\odot=0.30\,{\rm GeV\,cm^{-3}}$, consistent with local estimates \citep{Zhang_2013} and 9\% below the $0.33\,{\rm GeV\,cm^{-3}}$ of the OGLE halo \citep{Cautun2020,Mroz2024Nature}, and $q_h=0.72$, at the $1\sigma$ lower edge of the \citet{Ibata2024} range. At fixed $\rho_\odot$, this flattening alone lowers the MW dark-matter column toward the LMC by $11.8\%$ and the detected MW-halo PBH yield by $19\%$ (Table~\ref{tab:mw-flattening}). These column and detected-yield changes concern different observables from the halo optical depth, for which \citet{Mroz2025Reply} report few-percent changes among alternative Gaia DR3 halo models. With our baryonic components and $R_\odot=8.178\,{\rm kpc}$, the benchmark gives $V_c(27.31\,{\rm kpc})=179.1\,{\rm km\,s^{-1}}$, within the quoted uncertainties of the outermost measurement of \citet{Ou2024}, $173.0^{+15.8}_{-17.1}\,{\rm km\,s^{-1}}$, and $V_c(R_\odot)=215.2\,{\rm km\,s^{-1}}$. Because we compute circular speeds from the adopted gravitational potential, we also evaluate the potential of the reference configuration, combining the Cautun halo and nonstellar components with the Han--Gould disks used as lenses by \citet{Mroz2024Nature}. At $R_\odot$ this reconstruction gives $V_c\simeq215\,{\rm km\,s^{-1}}$, close to our benchmark. This is a comparison of the potentials generated by the mass components: \citet{Mroz2024Nature} retain the original fitted Cautun rotation curve when adopting the less massive Han--Gould disks. The benchmark nevertheless remains below the local measurement of \citet{Eilers2019}: at their adopted solar radius of $8.122\,{\rm kpc}$, our model gives $215.4\,{\rm km\,s^{-1}}$, compared with their $229.0\,{\rm km\,s^{-1}}$. 

Our \emph{stress test} takes $\rho_\odot=0.20\,{\rm GeV\,cm^{-3}}$, within the quoted uncertainties of \citet{Zhang_2013} and of the southern subsample of \citet{Guo_2020}, and $q_h=0.60$, deliberately beyond the values favored by \citet{Ibata2024}. It gives $V_c(R_\odot)=200.1\,{\rm km\,s^{-1}}$ and $V_c(50\,{\rm kpc})=132.9\,{\rm km\,s^{-1}}$. The latter matches extrapolations of Gaia DR3 rotation-curve models \citep{Ou2024,GarciaBellidoHawkins2024} but lies well below the $\simeq180\,{\rm km\,s^{-1}}$ implied by the enclosed mass measured from distant halo tracers with full 6D kinematics \citep{Medina2025}. The stress test is therefore a deliberately extreme case.

\emph{Disequilibrium.} The LMC drives a local wake, a large-scale collective response and a reflex displacement of the inner Galaxy \citep{GaravitoCamargo2019,GaravitoCamargo2021,Conroy2021,PetersenPenarrubia2021}. Its effect on the foreground lenses is set by the DM density along the sightline, which we evaluate with the basis-function expansion of the live MW--LMC $N$-body model of \citet{Lilleengen2023}, computed with EXP \citep{Petersen2022EXP}. Define
\begin{equation}
\kappa_{\rm w}(s)\equiv\frac{\rho_{\rm DM}(s)}{\rho^{\rm sph}_{\rm DM}(s)},
\label{eq:kappaw}
\end{equation}
with $s$ the heliocentric distance toward the LMC and $\rho^{\rm sph}_{\rm DM}$ the monopole of the present-day halo. Along the sightline $1.06\le\kappa_{\rm w}\le1.26$ (Fig.~\ref{fig:kappa}). Weighted by the optical-depth kernel and renormalized to the local density, as appropriate when $\rho_\odot$ is fixed by local measurements, the effective enhancement is $\simeq6\%$ relative to the present monopole and $\simeq12\%$ relative to the pre-infall halo. It comes from the dipole: the halo is denser on the LMC side. Only $\approx7\%$ of the MW-halo optical depth toward the LMC arises beyond $30$~kpc from the Galactic center, where the observed stellar wake lies \citep{Conroy2021}. LMC-induced disequilibrium therefore raises the expected MW-halo PBH signal; we nonetheless set $\kappa_{\rm w}=1$, which is conservative by $\simeq6\%$ (\ref{sec:kappa_sightline}).

\emph{LMC halo.} The benchmark replaces the Hernquist LMC halo with the cored Burkert profile fitted to the inner LMC rotation curve by \citet{Regis_2021}. Its enclosed mass within $13.2$~kpc, $2.43\times10^{10}\,M_\odot$, is consistent with the dynamical measurement of \citet{Watkins_2024} once stars and gas are added. The stress test uses the cored isothermal fit of the same work, truncated at $15$~kpc (see \ref{sec:lmc_dm}). Both cores lower the DM density where the source stars lie and hence the LMC-halo PBH rate. LMC-halo PBHs share the MW value of $f_{\rm PBH}$, with velocity dispersion set by each profile's enclosed mass.

\emph{Stellar lenses.} For the MW we adopt the thin and thick disks of \citet{Ibata2024}, which are more massive than the \citet{HanGould2003} disks of the reference calculation and bring $V_c(R_\odot)$ closer to measured values. We retain the circular-streaming prescription and isotropic $30\,{\rm km\,s^{-1}}$ dispersion of \citet{Mroz2024Nature}, with the circular speed computed from our adopted potential. For the LMC we retain the reference disk-plus-bar geometry and lower the stellar mass from $2.7$ to $2.1\times10^9\,M_\odot$, i.e., $M_\star/L_V=0.7$, at the lower end of the $0.9\pm0.2$ estimate of \citet{vanderMarel2002}. Stellar masses follow a \citet{Kroupa2001} broken-power-law initial mass function (IMF) with shallower slopes, $(\alpha_0,\alpha_1,\alpha_2,\alpha_3)=(-0.40,0.88,2.11,1.85)$: for $\alpha_1$--$\alpha_3$, the one-sided 95\% lower endpoints of each published range taken separately, and for $\alpha_0$, whose quoted interval has no calibrated confidence level, its lower endpoint (see \ref{sec:rate}). The combination is not a joint confidence region.

\paragraph{Rate and likelihood.}
A lens of mass $M$ at distance $D_L$ in front of a source at $D_S$ has $t_E=\theta_E/\mu$, with $\theta_E^2=(4GM/c^2)(D_S-D_L)/(D_LD_S)$ and $\mu$ the relative proper motion. The differential rate, $d^4\Gamma/(dD_L\,dM\,d^2\boldsymbol\mu)=2\theta_E\mu D_L^2\,(dn/dM)\,f(\boldsymbol\mu\,|\,D_L,D_S)$, is averaged over the LMC source distribution and projected onto $t_E$ (see~\ref{sec:rate}). PBHs have $dn_{\rm PBH}/dM=f_{\rm PBH}\,\rho_{\rm DM}\,\delta(M-M_{\rm PBH})/M_{\rm PBH}$. The expected number of events is
\begin{equation}
N_{\rm exp}=\sum_j N_{s,j}\,\Delta T_j\int\epsilon_j(t_E)\,\frac{d\Gamma_j}{dt_E}\,dt_E,
\label{eq:nexp}
\end{equation}
generalized to $\epsilon_j(t_E,\rho)$ for the high-cadence sample, where finite-source effects matter. Following the relaxed analysis of \citet{Mroz2024Nature} [their Eqs.~(1), (2) and (6)], we use the extended-Poisson likelihood
\begin{equation}
\begin{split}
\mathcal{L}(f,M)=\prod_{j}\Bigg[e^{-N_{{\rm exp},j}}\prod_{k=1}^{N_{{\rm obs},j}}N_{s,j}\Delta T_j\\
\times\frac1n\sum_{i=1}^n\epsilon_j(t_{E,k,i})\frac{d\Gamma_j}{dt_E}(t_{E,k,i})\Bigg],
\end{split}
\label{eq:likelihood}
\end{equation}
where $j$ labels field and observing phase, $k$ the detected events, and $i$ the $n=2\times10^5$ published posterior samples of each event's $t_E$. The rate sums all lens populations,
\begin{equation}
\frac{d\Gamma_j}{dt_E}=f_{\rm PBH}\left(\frac{d\Gamma^{\rm MW,h}_j}{dt_E}+\frac{d\Gamma^{\rm LMC,h}_j}{dt_E}\right)+\frac{d\Gamma^{\rm MW,\star}_j}{dt_E}+\frac{d\Gamma^{\rm LMC,\star}_j}{dt_E},
\label{eq:ratesum}
\end{equation}
with halo terms evaluated at $f_{\rm PBH}=1$. With a uniform prior on $f_{\rm PBH}\in[0,10]$, as in \citet{Mroz2024Nature}, we sample the posterior by Markov chain Monte Carlo in $\log_{10}f_{\rm PBH}$, including the Jacobian $f_{\rm PBH}\ln10$ so that the prior remains uniform in $f_{\rm PBH}$, and quote the 95th percentile of $f_{\rm PBH}$. Each detected event is thus shared between stellar and PBH lenses in proportion to their rates at its measured timescale, while the exponential penalizes the total expected count.

\begin{figure*}[t]
\centering
\includegraphics[width=0.8\textwidth]{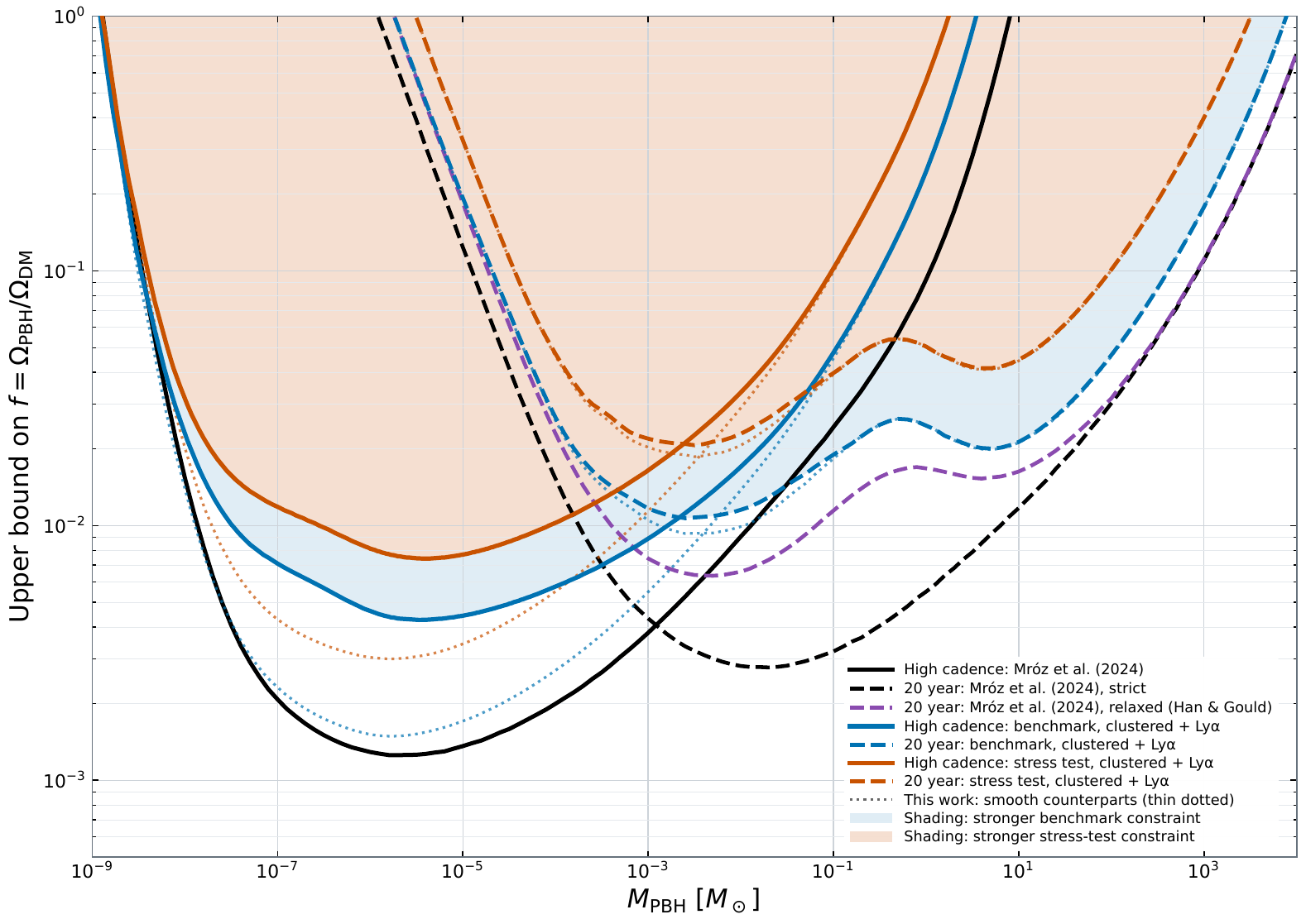}
\caption{Smooth-population 95\% upper limits and conditional clustered-PBH envelopes for the benchmark (blue) and stress test (orange). Thick colored curves: envelopes for PBHs in clusters of radius $R_{\rm cl}=30$~pc, subject to a Ly$\alpha$ threshold on $f_{\rm PBH}M_{\rm cl}$ (\ref{sec:clustered-pbhs}), for the high-cadence (solid) and 20-year OGLE-III$+$IV (dashed) surveys. Thin dotted colored curves: the corresponding limits for unclustered PBHs, to which the factors quoted in the text refer. Black: published high-cadence (solid) \citep{Mroz2024ApJL} and strict 20-year (dashed) \citep{Mroz2024Nature} limits; purple dashed: published relaxed 20-year limit for the Han--Gould disk \citep{Mroz2024Nature}, our reference curve. Shading marks, for each model, the region excluded by the stronger of its two clustered envelopes.}
\label{fig:constraints}
\end{figure*}

\paragraph{Results.}
Our full survey simulation predicts $2.86$ detected LMC self-lensing events in both models and $8.78$ ($8.64$) MW-disk events in the benchmark (stress test), compared with $5.7$ and $7.0$ in the reference calculation \citep{Mroz2024Nature}. The LMC prediction is halved, mainly by the lower stellar mass and the IMF, while the MW prediction rises by $25\%$ ($23\%$). The more massive Ibata disks alone raise the detected MW count by $\simeq30\%$ at fixed kinematics and IMF (\ref{sec:flattening}). The shallower IMF lowers the intrinsic stellar rate by $\simeq30\%$ at fixed mass density, but shifts events to longer, more detectable timescales, so detected counts fall by only $18$--$22\%$.

Figure~\ref{fig:constraints} shows the resulting limits. For the smooth 20-year calculation, over
$10^{-4}\leq M_{\rm PBH}/M_\odot\leq10$, the benchmark and stress test weaken the relaxed Han--Gould limit by maximum factors of $\simeq1.6$ and $\simeq3.4$, near $M_{\rm PBH}=0.32\,M_\odot$ and $0.05\,M_\odot$, respectively. Relative to the strict limit, the factors reach $6.2$ and $13.2$ near $0.25\,M_\odot$ and $0.20\,M_\odot$, respectively; this comparison also includes the event-origin difference quantified by OGLE. The weakening is mass dependent because the MW halo, LMC halo and stellar populations reshape the efficiency-weighted timescale distribution differently at different masses, and the measured timescales determine which PBH masses can absorb the detections; no counting-only rescaling reproduces it. At $M_{\rm PBH}=0.45\,M_\odot$, a typical stellar-lens mass at which PBH and stellar timescales overlap and every ingredient acts, restoring the canonical IMF alone tightens the benchmark limit by a factor $1.46$, the largest individual effect, while restoring all halo or all stellar ingredients gives comparable tightening factors of $1.32$ and $1.29$, respectively (Table~\ref{tab:ablation}).

For comparison with \citet{Mroz2024Nature}, we evaluate the likelihood on their 101-by-101 grid in mass and abundance. Both models attain their grid maxima at $M_{\rm best}=0.0316\,M_\odot$. Defining $\Delta\chi^2\equiv2\ln[\mathcal L_{\max}/\mathcal L(f=10^{-4},M_{\rm best})]$, we obtain $9.15$ and $10.23$ for the benchmark and stress test, respectively, compared with the $3.7$ reported for their Han--Gould model. Replacing the denominator by $\max_M\mathcal L(f=10^{-4},M)$ gives $8.30$ and $9.57$. Thus, the weaker upper limits coexist with larger conditional likelihood improvements than reported for the Han--Gould model. 

For the high-cadence survey, we include MW-halo PBH lensing toward both the LMC and SMC, together with the LMC-halo contribution, and neglect the SMC-halo contribution as in \cite{Mroz2024ApJL}. We compare at $M_{\rm PBH}=2.00\times10^{-6}\,M_\odot$, the tabulated minimum of the published curve, where $f_{95}^{\rm ref}=1.256\times10^{-3}$. Here the single detected event, with $t_E\simeq90$~d, carries negligible PBH likelihood, so $f_{95}\simeq3.0/N_{\rm exp}^{\rm PBH}(f_{\rm PBH}=1)$ and the limit tracks the halo column directly. We obtain $f_{95}=1.49\times10^{-3}$ and $3.00\times10^{-3}$, weaker by factors $1.19$ and $2.39$.

\paragraph{Clustered PBHs.}
If PBHs formed in clusters, the counts become super-Poissonian and the planetary-mass limits weaken further, within bounds set by the Ly$\alpha$ forest on cluster shot noise \citep{De_Luca_2022,IvanovTrifinopoulos2026}. For clusters of radius $30$~pc and the adopted threshold $fM_{\rm cl}\leq8.6\times10^2\,M_\odot$, the high-cadence conditional envelopes at $M_{\rm PBH}=2.00\times10^{-6}\,M_\odot$ are weaker than the published limit by factors $\simeq3.44$ and $\simeq6.03$ for the benchmark and stress test, respectively (see~\ref{sec:clustered-pbhs}). For the 20-year sample, clustering weakens the benchmark and stress-test limits relative to the corresponding smooth calculations by at most $16.4\%$ and $11.4\%$, near $M_{\rm PBH}=5.0\times10^{-3}$ and $6.3\times10^{-3}\,M_\odot$, respectively. Figure~\ref{fig:constraints} shows the clustered envelopes and their smooth counterparts for both surveys. These envelopes impose the microlensing and Ly$\alpha$ thresholds separately and are not joint posterior limits.

\paragraph{Discussion.}
At fixed data, inputs that are each motivated by recent measurements move OGLE's bound on compact-object DM by factors of $\sim1.2$--$1.6$, and a deliberately extreme configuration moves it by up to $\sim3.4$. Because our event-origin treatment matches OGLE's relaxed analysis, this weakening reflects the rate model, not a reinterpretation of the detections. One effect often invoked to weaken the bound, LMC-induced halo disequilibrium, goes the other way. Clustering acts differently in the two windows: it weakens the planetary-mass bound by up to a factor $\simeq3.4$--$6.0$ relative to the published limit, bounded mainly by the Ly$\alpha$ forest, but moves the 20-year bound by at most $16\%$, so it bears on the planetary-mass interpretation of the bulge events rather than on the stellar-mass window. The planetary-mass PBH interpretation tested by the high-cadence survey concerns OGLE bulge short-timescale events \citep{Niikura2019Bulge}. We do not claim that our benchmark restores compatibility with that population, with the anomalous bulge-lens population of \citet{Perkins2025}, or with the larger abundances argued by \citet{Hawkins2025}; each requires a matched mass-function and survey comparison.

Five caveats bound our conclusions. First, halo parameters are fixed, not marginalized, and our values of $\rho_\odot$, $q_h$ and the LMC core were selected to suppress the rate; a joint treatment of halo uncertainties and their correlations is needed for a calibrated bound. Second, we use monochromatic PBH masses; an extended mass function changes both the signal and the timescale likelihood and is not guaranteed to weaken the bound. Third, we do not reanalyze the long-timescale SMC sample \citep{Mroz2025ApJS}. Fourth, field-specific source luminosity functions were available to us for only one LMC and one SMC field (see \ref{sec:lf_inputs}). Fifth, the clustered envelopes impose the microlensing and Ly$\alpha$ thresholds separately and assume primordially formed, independently distributed clusters; a stream budget at the combined-stream value would reduce both envelopes to a factor $\simeq2.4$ relative to the published limit (Table~\ref{tab:cluster-budget-results}).

A joint hierarchical analysis of the OGLE bulge and Magellanic Cloud samples, with a common halo-kinematic and PBH mass-function prior, would confront the bulge preference for a PBH-like component \citep{Perkins2025} with the absence of identified PBH events toward the Magellanic Clouds in one framework. Until such analyses exist, applications of microlensing bounds on compact-object DM should state the halo and stellar-population assumptions behind the curve they use, and whether it is a strict or a relaxed limit. Where robustness to those assumptions matters, the benchmark limits presented here are a conservative alternative to the published curves.\\

\begin{acknowledgments}
We thank P.~Mr\'oz for providing detection efficiencies, source counts, timescale posteriors and source luminosity functions and for helpful clarifications of his methodology.  This work is partly supported by the U.S.\ Department of Energy grant number DE-SC0010107 (SP).
\end{acknowledgments}

\bibliographystyle{apsrev4-2}
\bibliography{biblio}

@article{Bienayme2024,
  author        = {Bienaym{\'e}, O. and Robin, A. C. and Salomon, J.-B. and Reyl{\'e}, C.},
  title         = {{Dark matter in the Milky Way: Measurements up to 3 kpc from the Galactic plane above the Sun}},
  journal       = {Astron. Astrophys.},
  year          = {2024},
  volume        = {689},
  pages         = {A280},
  doi           = {10.1051/0004-6361/202450327},
  eprint        = {2406.08158},
  archivePrefix = {arXiv}
}

@article{Calcino2018,
  author        = {Calcino, Josh and Garc{\'i}a-Bellido, Juan and Davis, Tamara M.},
  title         = {{Updating the MACHO fraction of the Milky Way dark halo with improved mass models}},
  journal       = {Mon. Not. R. Astron. Soc.},
  year          = {2018},
  volume        = {479},
  pages         = {2889},
  doi           = {10.1093/mnras/sty1368},
  eprint        = {1803.09205},
  archivePrefix = {arXiv}
}

@article{CalchiNovati2009,
  author        = {Calchi Novati, S. and Mancini, L. and Scarpetta, G. and Wyrzykowski, {\L}.},
  title         = {{Large Magellanic Cloud self-lensing for OGLE-II microlensing observations}},
  journal       = {Mon. Not. R. Astron. Soc.},
  year          = {2009},
  volume        = {400},
  pages         = {1625--1631},
  doi           = {10.1111/j.1365-2966.2009.15570.x}
}

@article{Cautun2020,
  author        = {Cautun, Marius and Ben{\'i}tez-Llambay, Alejandro and Deason, Alis J. and Frenk, Carlos S. and Fattahi, Azadeh and G{\'o}mez, Facundo A. and Grand, Robert J. J. and Oman, Kyle A. and Navarro, Julio F. and Simpson, Christine M.},
  title         = {{The Milky Way total mass profile as inferred from Gaia DR2}},
  journal       = {Mon. Not. R. Astron. Soc.},
  year          = {2020},
  volume        = {494},
  pages         = {4291--4313},
  doi           = {10.1093/mnras/staa1017},
  eprint        = {1911.04557},
  archivePrefix = {arXiv}
}

@article{Conroy2021,
  author        = {Conroy, Charlie and Naidu, Rohan P. and Garavito-Camargo, Nicol{\'a}s and Besla, Gurtina and Zaritsky, Dennis and Bonaca, Ana and Johnson, Benjamin D.},
  title         = {{All-sky dynamical response of the Galactic halo to the Large Magellanic Cloud}},
  journal       = {Nature},
  year          = {2021},
  volume        = {592},
  pages         = {534--536},
  doi           = {10.1038/s41586-021-03385-7},
  eprint        = {2104.09515},
  archivePrefix = {arXiv}
}

@article{Erkal2019,
  author        = {Erkal, D. and Belokurov, V. and Laporte, C. F. P. and Koposov, S. E. and Li, T. S. and Grillmair, C. J. and Kallivayalil, N. and Price-Whelan, A. M. and Evans, N. W. and Hawkins, K. and Hendel, D. and Mateu, C. and Navarro, J. F. and del Pino, A. and Slater, C. T. and Sohn, S. T.},
  collaboration = {OATs: Orphan Aspen Treasury Collaboration},
  title         = {{The total mass of the Large Magellanic Cloud from its perturbation on the Orphan stream}},
  journal       = {Mon. Not. R. Astron. Soc.},
  year          = {2019},
  volume        = {487},
  pages         = {2685--2700},
  doi           = {10.1093/mnras/stz1371},
  eprint        = {1812.08192},
  archivePrefix = {arXiv}
}

@article{EvansBowden2014,
  author        = {Evans, N. W. and Bowden, A.},
  title         = {{Extremely flat haloes and the shape of the Galaxy}},
  journal       = {Mon. Not. R. Astron. Soc.},
  year          = {2014},
  volume        = {443},
  pages         = {2--11},
  doi           = {10.1093/mnras/stu1113},
  eprint        = {1406.3710},
  archivePrefix = {arXiv}
}

@article{Fushimi2024,
  author        = {Fushimi, K. J. and Mosquera, M. E. and Dominguez, M.},
  title         = {{A determination of the Large Magellanic Cloud dark matter subhalo mass using the Milky Way halo stars in its gravitational wake}},
  journal       = {Astron. Astrophys.},
  year          = {2024},
  volume        = {688},
  pages         = {A147},
  doi           = {10.1051/0004-6361/202347450},
  eprint        = {2309.12989},
  archivePrefix = {arXiv}
}

@article{GaravitoCamargo2019,
  author        = {Garavito-Camargo, Nicol{\'a}s and Besla, Gurtina and Laporte, Chervin F. P. and Johnston, Kathryn V. and G{\'o}mez, Facundo A. and Watkins, Laura L.},
  title         = {{Hunting for the Dark Matter Wake Induced by the Large Magellanic Cloud}},
  journal       = {Astrophys. J.},
  year          = {2019},
  volume        = {884},
  pages         = {51},
  doi           = {10.3847/1538-4357/ab32eb}
}

@article{GaravitoCamargo2021,
  author        = {Garavito-Camargo, Nicol{\'a}s and Besla, Gurtina and Laporte, Chervin F. P. and Price-Whelan, Adrian M. and Cunningham, Emily C. and Johnston, Kathryn V. and Weinberg, Martin and G{\'o}mez, Facundo A.},
  title         = {{Quantifying the Impact of the Large Magellanic Cloud on the Structure of the Milky Way's Dark Matter Halo Using Basis Function Expansions}},
  journal       = {Astrophys. J.},
  year          = {2021},
  volume        = {919},
  pages         = {109},
  doi           = {10.3847/1538-4357/ac0b44},
  eprint        = {2010.00816},
  archivePrefix = {arXiv}
}

@article{Gyuk2000,
  author        = {Gyuk, G. and Dalal, N. and Griest, K.},
  title         = {{Self-lensing Models of the Large Magellanic Cloud}},
  journal       = {Astrophys. J.},
  year          = {2000},
  volume        = {535},
  pages         = {90--103},
  doi           = {10.1086/308834}
}

@article{HanGould2003,
  author        = {Han, Cheongho and Gould, Andrew},
  title         = {{Stellar Contribution to the Galactic Bulge Microlensing Optical Depth}},
  journal       = {Astrophys. J.},
  year          = {2003},
  volume        = {592},
  pages         = {172--175},
  doi           = {10.1086/375706},
  eprint        = {astro-ph/0303309},
  archivePrefix = {arXiv}
}

@article{Hawkins2025,
  author        = {Hawkins, M. R. S. and Garc{\'i}a-Bellido, J.},
  title         = {{A critical analysis of the recent OGLE limits on stellar mass primordial black holes in the halo of the Milky Way}},
  journal       = {Mon. Not. R. Astron. Soc.},
  year          = {2025},
  volume        = {544},
  pages         = {1950--1957},
  doi           = {10.1093/mnras/staf1826},
  eprint        = {2509.05400},
  archivePrefix = {arXiv}
}

@article{Ibata2024,
  author        = {Ibata, Rodrigo and Malhan, Khyati and Tenachi, Wassim and Ardern-Arentsen, Anke and Bellazzini, Michele and Bianchini, Paolo and Bonifacio, Piercarlo and Caffau, Elisabetta and Diakogiannis, Foivos and Errani, Raphael and Famaey, Benoit and Ferrone, Salvatore and Martin, Nicolas F. and di Matteo, Paola and Monari, Giacomo and Renaud, Florent and Starkenburg, Else and Thomas, Guillaume and Viswanathan, Akshara and Yuan, Zhen},
  title         = {{Charting the Galactic Acceleration Field. II. A Global Mass Model of the Milky Way from the STREAMFINDER Atlas of Stellar Streams Detected in Gaia DR3}},
  journal       = {Astrophys. J.},
  year          = {2024},
  volume        = {967},
  pages         = {89},
  doi           = {10.3847/1538-4357/ad382d},
  eprint        = {2311.17202},
  archivePrefix = {arXiv}
}

@article{Kroupa2001,
  author        = {Kroupa, P.},
  title         = {{On the variation of the initial mass function}},
  journal       = {Mon. Not. R. Astron. Soc.},
  year          = {2001},
  volume        = {322},
  pages         = {231--246},
  doi           = {10.1046/j.1365-8711.2001.04022.x},
  eprint        = {astro-ph/0009005},
  archivePrefix = {arXiv}
}

@article{Lilleengen2023,
  author        = {Lilleengen, Sophia and Petersen, Michael S. and Erkal, Denis and Pe{\~n}arrubia, Jorge and Koposov, Sergey E. and Li, Ting S. and Cullinane, Lara R. and Ji, Alexander P. and Kuehn, Kyler and Lewis, Geraint F. and Mackey, Dougal and Pace, Andrew B. and Shipp, Nora and Zucker, Daniel B. and Bland-Hawthorn, Joss and Hilmi, Tariq},
  collaboration = {S5 Collaboration},
  title         = {{The effect of the deforming dark matter haloes of the Milky Way and the Large Magellanic Cloud on the Orphan--Chenab stream}},
  journal       = {Mon. Not. R. Astron. Soc.},
  year          = {2023},
  volume        = {518},
  pages         = {774--790},
  doi           = {10.1093/mnras/stac3108},
  eprint        = {2205.01688},
  archivePrefix = {arXiv}
}

@article{Mancini2004,
  author        = {Mancini, L. and Calchi Novati, S. and Jetzer, Ph. and Scarpetta, G.},
  title         = {{LMC self-lensing from a new perspective}},
  journal       = {Astron. Astrophys.},
  year          = {2004},
  volume        = {427},
  pages         = {61--77},
  doi           = {10.1051/0004-6361:20034535},
  eprint        = {astro-ph/0405257},
  archivePrefix = {arXiv}
}

@article{McMillan2017,
  author        = {McMillan, Paul J.},
  title         = {{The mass distribution and gravitational potential of the Milky Way}},
  journal       = {Mon. Not. R. Astron. Soc.},
  year          = {2017},
  volume        = {465},
  pages         = {76--94},
  doi           = {10.1093/mnras/stw2759},
  eprint        = {1608.00971},
  archivePrefix = {arXiv}
}

@article{Mroz2024ApJL,
  author        = {Mr{\'o}z, Przemek and Udalski, Andrzej and Szyma{\'n}ski, Micha{\l} K. and Soszy{\'n}ski, Igor and Pietrukowicz, Pawe{\l} and Koz{\l}owski, Szymon and Poleski, Rados{\l}aw and Skowron, Jan and Ulaczyk, Krzysztof and Gromadzki, Mariusz and Rybicki, Krzysztof and Iwanek, Patryk and Wrona, Marcin and Mr{\'o}z, Mateusz J.},
  title         = {{Limits on Planetary-mass Primordial Black Holes from the OGLE High-cadence Survey of the Magellanic Clouds}},
  journal       = {Astrophys. J. Lett.},
  year          = {2024},
  volume        = {976},
  pages         = {L19},
  doi           = {10.3847/2041-8213/ad8e68},
  eprint        = {2410.06251},
  archivePrefix = {arXiv}
}

@article{Mroz2024ApJS,
  author        = {Mr{\'o}z, Przemek and Udalski, Andrzej and Szyma{\'n}ski, Micha{\l} K. and Kapusta, Mateusz and Soszy{\'n}ski, Igor and Wyrzykowski, {\L}ukasz and Pietrukowicz, Pawe{\l} and Koz{\l}owski, Szymon and Poleski, Rados{\l}aw and Skowron, Jan and Skowron, Dorota and Ulaczyk, Krzysztof and Gromadzki, Mariusz and Rybicki, Krzysztof and Iwanek, Patryk and Wrona, Marcin and Ratajczak, Milena},
  title         = {{Microlensing Optical Depth and Event Rate toward the Large Magellanic Cloud Based on 20 yr of OGLE Observations}},
  journal       = {Astrophys. J. Suppl. Ser.},
  year          = {2024},
  volume        = {273},
  pages         = {4},
  doi           = {10.3847/1538-4365/ad452e},
  eprint        = {2403.02398},
  archivePrefix = {arXiv}
}

@article{Mroz2024Nature,
  author        = {Mr{\'o}z, Przemek and Udalski, Andrzej and Szyma{\'n}ski, Micha{\l} K. and Soszy{\'n}ski, Igor and Wyrzykowski, {\L}ukasz and Pietrukowicz, Pawe{\l} and Koz{\l}owski, Szymon and Poleski, Rados{\l}aw and Skowron, Jan and Skowron, Dorota and Ulaczyk, Krzysztof and Gromadzki, Mariusz and Rybicki, Krzysztof and Iwanek, Patryk and Wrona, Marcin and Ratajczak, Milena},
  title         = {{No massive black holes in the Milky Way halo}},
  journal       = {Nature},
  year          = {2024},
  volume        = {632},
  pages         = {749--751},
  doi           = {10.1038/s41586-024-07704-6},
  eprint        = {2403.02386},
  archivePrefix = {arXiv}
}

@article{Mroz2025ApJS,
  author        = {Mr{\'o}z, Przemek and Udalski, Andrzej and Szyma{\'n}ski, Micha{\l} K. and Soszy{\'n}ski, Igor and Pietrukowicz, Pawe{\l} and Koz{\l}owski, Szymon and Poleski, Rados{\l}aw and Skowron, Jan and Skowron, Dorota and Ulaczyk, Krzysztof and Gromadzki, Mariusz and Rybicki, Krzysztof and Iwanek, Patryk and Wrona, Marcin and Ratajczak, Milena},
  title         = {{Microlensing Optical Depth, Event Rate, and Limits on Compact Objects in Dark Matter Based on 20 Yr of OGLE Observations of the Small Magellanic Cloud}},
  journal       = {Astrophys. J. Suppl. Ser.},
  year          = {2025},
  volume        = {280},
  pages         = {49},
  doi           = {10.3847/1538-4365/adf842},
  eprint        = {2507.13794},
  archivePrefix = {arXiv}
}

@article{NibauerBonaca2025,
  author        = {Nibauer, Jacob and Bonaca, Ana},
  title         = {{Galactic Accelerations from the GD-1 Stream Suggest a Tilted Dark Matter Halo}},
  journal       = {Astrophys. J. Lett.},
  year          = {2025},
  volume        = {985},
  pages         = {L22},
  doi           = {10.3847/2041-8213/add0a9},
  eprint        = {2504.07187},
  archivePrefix = {arXiv}
}

@article{Niikura2019Bulge,
  author        = {Niikura, Hiroko and Takada, Masahiro and Yokoyama, Shuichiro and Sumi, Takahiro and Masaki, Shogo},
  title         = {{Constraints on Earth-mass primordial black holes from OGLE 5-year microlensing events}},
  journal       = {Phys. Rev. D},
  year          = {2019},
  volume        = {99},
  pages         = {083503},
  doi           = {10.1103/PhysRevD.99.083503},
  eprint        = {1901.07120},
  archivePrefix = {arXiv}
}

@article{Ou2024,
  author        = {Ou, Xiaowei and Eilers, Anna-Christina and Necib, Lina and Frebel, Anna},
  title         = {{The dark matter profile of the Milky Way inferred from its circular velocity curve}},
  journal       = {Mon. Not. R. Astron. Soc.},
  year          = {2024},
  volume        = {528},
  pages         = {693--710},
  doi           = {10.1093/mnras/stae034},
  eprint        = {2303.12838},
  archivePrefix = {arXiv}
}

@misc{Perkins2025,
  author        = {Perkins, Scott E. and McGill, Peter and Dawson, William A. and Ho, Ming-Feng and Abrams, Natasha S. and Bird, Simeon and Lu, Jessica R.},
  title         = {{Hints of an Anomalous Lens Population towards the Galactic Bulge}},
  year          = {2025},
  eprint        = {2503.22037},
  archivePrefix = {arXiv},
  primaryClass  = {astro-ph.GA}
}

@article{Petersen2022EXP,
  author        = {Petersen, Michael S. and Weinberg, Martin D. and Katz, Neal},
  title         = {{EXP: N-body integration using basis function expansions}},
  journal       = {Mon. Not. R. Astron. Soc.},
  year          = {2022},
  volume        = {510},
  pages         = {6201--6217},
  doi           = {10.1093/mnras/stab3639}
}

@article{PetersenPenarrubia2021,
  author        = {Petersen, Michael S. and Pe{\~n}arrubia, Jorge},
  title         = {{Detection of the Milky Way reflex motion due to the Large Magellanic Cloud infall}},
  journal       = {Nat. Astron.},
  year          = {2021},
  volume        = {5},
  pages         = {251--255},
  doi           = {10.1038/s41550-020-01254-3},
  eprint        = {2011.10581},
  archivePrefix = {arXiv}
}

@article{Tisserand2007,
  author        = {Tisserand, P. and Le Guillou, L. and Afonso, C. and Albert, J. N. and Andersen, J. and Ansari, R. and Aubourg, {\'E}. and Bareyre, P. and Beaulieu, J. P. and Charlot, X. and Coutures, C. and Ferlet, R. and Fouqu{\'e}, P. and Glicenstein, J. F. and Goldman, B. and Gould, A. and Graff, D. and Gros, M. and Haissinski, J. and Hamadache, C. and de Kat, J. and Lasserre, T. and Lesquoy, {\'E}. and Loup, C. and Magneville, C. and Marquette, J. B. and Maurice, {\'E}. and Maury, A. and Milsztajn, A. and Moniez, M. and Palanque-Delabrouille, N. and Perdereau, O. and Rahal, Y. R. and Rich, J. and Spiro, M. and Vidal-Madjar, A. and Vigroux, L. and Zylberajch, S.},
  collaboration = {EROS-2 Collaboration},
  title         = {{Limits on the Macho content of the Galactic Halo from the EROS-2 Survey of the Magellanic Clouds}},
  journal       = {Astron. Astrophys.},
  year          = {2007},
  volume        = {469},
  pages         = {387--404},
  doi           = {10.1051/0004-6361:20066017}
}

@article{vanderMarel2002,
  author        = {van der Marel, Roeland P. and Alves, David R. and Hardy, Eduardo and Suntzeff, Nicholas B.},
  title         = {{New Understanding of Large Magellanic Cloud Structure, Dynamics, and Orbit from Carbon Star Kinematics}},
  journal       = {Astron. J.},
  year          = {2002},
  volume        = {124},
  pages         = {2639--2663},
  doi           = {10.1086/343775},
  eprint        = {astro-ph/0205161},
  archivePrefix = {arXiv}
}

@article{Wyrzykowski2011LMC,
  author        = {Wyrzykowski, {\L}. and Koz{\l}owski, S. and Skowron, J. and Udalski, A. and Szyma{\'n}ski, M. K. and Kubiak, M. and Pietrzy{\'n}ski, G. and Soszy{\'n}ski, I. and Szewczyk, O. and Ulaczyk, K. and Poleski, R.},
  title         = {{The OGLE view of microlensing towards the Magellanic Clouds -- III. Ruling out subsolar MACHOs with the OGLE-III LMC data}},
  journal       = {Mon. Not. R. Astron. Soc.},
  year          = {2011},
  volume        = {413},
  pages         = {493--508},
  doi           = {10.1111/j.1365-2966.2010.18150.x},
  eprint        = {1012.1154},
  archivePrefix = {arXiv}
}

@article{Wyrzykowski2011SMC,
  author        = {Wyrzykowski, {\L}. and Skowron, J. and Koz{\l}owski, S. and Udalski, A. and Szyma{\'n}ski, M. K. and Kubiak, M. and Pietrzy{\'n}ski, G. and Soszy{\'n}ski, I. and Szewczyk, O. and Ulaczyk, K. and Poleski, R. and Tisserand, P.},
  title         = {{The OGLE view of microlensing towards the Magellanic Clouds -- IV. OGLE-III SMC data and final conclusions on MACHOs}},
  journal       = {Mon. Not. R. Astron. Soc.},
  year          = {2011},
  volume        = {416},
  pages         = {2949--2961},
  doi           = {10.1111/j.1365-2966.2011.19243.x},
  eprint        = {1106.2925},
  archivePrefix = {arXiv}
}

@article{Regis_2021,
  author  = {Regis, Marco and Reynoso-Cordova, Javier and Filipovi{\'c}, Miroslav D. and Br{\"u}ggen, Marcus and Carretti, Ettore and Collier, Jordan and Hopkins, Andrew M. and Lenc, Emil and Maio, Umberto and Marvil, Joshua R. and Norris, Ray P. and Vernstrom, Tessa},
  title   = {{The EMU view of the Large Magellanic Cloud: troubles for sub-TeV WIMPs}},
  journal = {J. Cosmol. Astropart. Phys.},
  year    = {2021},
  volume  = {2021},
  number  = {11},
  pages   = {046},
  doi     = {10.1088/1475-7516/2021/11/046}
}

@article{Watkins_2024,
  author        = {Watkins, Laura L. and van der Marel, Roeland P. and Bennet, Paul},
  title         = {{The Mass of the Large Magellanic Cloud from the Three-dimensional Kinematics of Its Globular Clusters}},
  journal       = {Astrophys. J.},
  year          = {2024},
  volume        = {963},
  pages         = {84},
  doi           = {10.3847/1538-4357/ad1f58},
  eprint        = {2401.14458},
  archivePrefix = {arXiv},
  primaryClass  = {astro-ph.GA}
}

@article{Guo_2020,
  author  = {Guo, Rui and Liu, Chao and Mao, Shude and Xue, Xiang-Xiang and Long, R. J. and Zhang, Lan},
  title   = {{Measuring the local dark matter density with LAMOST DR5 and Gaia DR2}},
  journal = {Mon. Not. R. Astron. Soc.},
  year    = {2020},
  volume  = {495},
  pages   = {4828--4844},
  doi     = {10.1093/mnras/staa1483}
}

@article{Zhang_2013,
  author  = {Zhang, Lan and Rix, Hans-Walter and van de Ven, Glenn and Bovy, Jo and Liu, Chao and Zhao, Gang},
  title   = {{The Gravitational Potential near the Sun from SEGUE K-dwarf Kinematics}},
  journal = {Astrophys. J.},
  year    = {2013},
  volume  = {772},
  pages   = {108},
  doi     = {10.1088/0004-637X/772/2/108}
}

@article{GarciaBellidoHawkins2024,
  author        = {Garc{\'i}a-Bellido, Juan and Hawkins, Michael},
  title         = {{Reanalysis of the MACHO Constraints on PBH in the Light of Gaia DR3 Data}},
  journal       = {Universe},
  volume        = {10},
  pages         = {449},
  year          = {2024},
  doi           = {10.3390/universe10120449},
  eprint        = {2402.00212},
  archivePrefix = {arXiv},
  primaryClass  = {astro-ph.GA}
}

@article{De_Luca_2022,
  author  = {De Luca, Valerio and Franciolini, Gabriele and Riotto, Antonio and Veerm{\"a}e, Hardi},
  title   = {{Ruling Out Initially Clustered Primordial Black Holes as Dark Matter}},
  journal = {Phys. Rev. Lett.},
  year    = {2022},
  volume  = {129},
  pages   = {191302},
  doi     = {10.1103/PhysRevLett.129.191302}
}

@article{IvanovTrifinopoulos2026,
  author        = {Ivanov, Mikhail M. and Trifinopoulos, Sokratis},
  title         = {{Effective Field Theory Constraints on Primordial Black Holes from the High-Redshift Lyman-$\alpha$ Forest}},
  journal       = {Phys. Rev. Lett.},
  volume        = {136},
  pages         = {171402},
  year          = {2026},
  doi           = {10.1103/8g8z-bmxd},
  eprint        = {2508.04767},
  archivePrefix = {arXiv},
  primaryClass  = {astro-ph.CO}
}

@article{Murgia_2019,
  author        = {Murgia, Riccardo and Scelfo, Giulio and Viel, Matteo and Raccanelli, Alvise},
  title         = {{Lyman-$\alpha$ Forest Constraints on Primordial Black Holes as Dark Matter}},
  journal       = {Phys. Rev. Lett.},
  year          = {2019},
  volume        = {123},
  pages         = {071102},
  doi           = {10.1103/PhysRevLett.123.071102},
  eprint        = {1903.10509},
  archivePrefix = {arXiv}
}

@article{Gorton_2022,
  author        = {Gorton, Matthew and Green, Anne M.},
  title         = {{Effect of clustering on primordial black hole microlensing constraints}},
  journal       = {J. Cosmol. Astropart. Phys.},
  year          = {2022},
  volume        = {2022},
  number        = {08},
  pages         = {035},
  doi           = {10.1088/1475-7516/2022/08/035},
  eprint        = {2203.04209},
  archivePrefix = {arXiv}
}

@article{Peta__2022,
  author        = {Peta{\v c}, Mihael and Lavalle, Julien and Jedamzik, Karsten},
  title         = {{Microlensing constraints on clustered primordial black holes}},
  journal       = {Phys. Rev. D},
  year          = {2022},
  volume        = {105},
  pages         = {083520},
  doi           = {10.1103/PhysRevD.105.083520},
  eprint        = {2201.02521},
  archivePrefix = {arXiv}
}

@article{Banik_2021,
  author        = {Banik, Nilanjan and Bovy, Jo and Bertone, Gianfranco and Erkal, Denis and de Boer, T. J. L.},
  title         = {{Evidence of a population of dark subhaloes from Gaia and Pan-STARRS observations of the GD-1 stream}},
  journal       = {Mon. Not. R. Astron. Soc.},
  year          = {2021},
  volume        = {502},
  pages         = {2364--2380},
  doi           = {10.1093/mnras/stab210},
  eprint        = {1911.02662},
  archivePrefix = {arXiv}
}

@article{NunoSilesGarciaBellido2025,
  author        = {Nu{\~n}o Siles, Jos{\'e} Francisco and Garc{\'i}a-Bellido, Juan},
  title         = {{Primordial black hole clusters, phenomenology \& implications}},
  journal       = {Phys. Dark Univ.},
  volume        = {47},
  pages         = {101789},
  year          = {2025},
  doi           = {10.1016/j.dark.2024.101789},
  eprint        = {2405.06391},
  archivePrefix = {arXiv},
  primaryClass  = {astro-ph.CO}
}

@article{Toshchenko2026,
  author        = {Toshchenko, K. A. and Baklanov, P. V. and Belotsky, K. M. and Blinnikov, S. I.},
  title         = {{Microlensing by a Cluster of Primordial Black Holes}},
  journal       = {Astron. Lett.},
  volume        = {51},
  pages         = {461--469},
  year          = {2025},
  doi           = {10.1134/S1063773725700562},
  eprint        = {2603.23412},
  archivePrefix = {arXiv},
  primaryClass  = {astro-ph.CO}
}

@article{Zeldovich1967,
  author  = {Zel'dovich, Ya. B. and Novikov, I. D.},
  title   = {The hypothesis of cores retarded during expansion and the hot cosmological model},
  journal = {Soviet Astronomy},
  volume  = {10},
  pages   = {602},
  year    = {1967}
}

@article{Hawking1971,
  author  = {Hawking, S.},
  title   = {Gravitationally collapsed objects of very low mass},
  journal = {Mon. Not. R. Astron. Soc.},
  volume  = {152},
  pages   = {75},
  year    = {1971}
}

@article{CarrHawking1974,
  author  = {Carr, B. J. and Hawking, S. W.},
  title   = {Black holes in the early Universe},
  journal = {Mon. Not. R. Astron. Soc.},
  volume  = {168},
  pages   = {399},
  year    = {1974}
}

@article{CarrKuhnel2020,
  author  = {Carr, Bernard and K{\"u}hnel, Florian},
  title   = {Primordial Black Holes as Dark Matter: Recent Developments},
  journal = {Ann. Rev. Nucl. Part. Sci.},
  volume  = {70},
  pages   = {355},
  year    = {2020},
  eprint  = {2006.02838},
  archivePrefix = {arXiv}
}

@article{Carr2021RPP,
  author  = {Carr, Bernard and Kohri, Kazunori and Sendouda, Yuuiti and Yokoyama, Jun'ichi},
  title   = {Constraints on primordial black holes},
  journal = {Rept. Prog. Phys.},
  volume  = {84},
  pages   = {116902},
  year    = {2021},
  eprint  = {2002.12778},
  archivePrefix = {arXiv}
}

@article{GreenKavanagh2021,
  author  = {Green, Anne M. and Kavanagh, Bradley J.},
  title   = {Primordial Black Holes as a dark matter candidate},
  journal = {J. Phys. G},
  volume  = {48},
  pages   = {043001},
  year    = {2021},
  eprint  = {2007.10722},
  archivePrefix = {arXiv}
}

@article{Paczynski1986,
  author  = {Paczy{\'n}ski, B.},
  title   = {Gravitational microlensing by the galactic halo},
  journal = {Astrophys. J.},
  volume  = {304},
  pages   = {1},
  year    = {1986}
}

@article{Mroz2025Reply,
  author        = {Mr{\'o}z, Przemek and Udalski, Andrzej and Szyma{\'n}ski, Micha{\l} K. and Soszy{\'n}ski, Igor and Wyrzykowski, {\L}ukasz and Pietrukowicz, Pawe{\l} and Koz{\l}owski, Szymon and Poleski, Rados{\l}aw and Skowron, Jan and Skowron, Dorota and Ulaczyk, Krzysztof and Gromadzki, Mariusz and Rybicki, Krzysztof and Iwanek, Patryk and Wrona, Marcin and Ratajczak, Milena},
  title         = {{Eppur non si trovano: Comments on the Primordial Black Hole Limits in the Galactic Halo}},
  journal       = {Acta Astron.},
  volume        = {75},
  pages         = {173--208},
  year          = {2025},
  doi           = {10.32023/0001-5237/75.3.1},
  eprint        = {2510.20005},
  archivePrefix = {arXiv}
}

@misc{Medina2025,
  author        = {Medina, Gustavo E. and Li, Ting S. and Eadie, Gwendolyn M. and others},
  title         = {{The mass of the Milky Way from outer halo stars measured by DESI DR1}},
  year          = {2025},
  eprint        = {2508.19351},
  archivePrefix = {arXiv},
  primaryClass  = {astro-ph.GA}
}

@article{Eilers2019,
  author  = {Eilers, Anna-Christina and Hogg, David W. and Rix, Hans-Walter and Ness, Melissa K.},
  title   = {{The Circular Velocity Curve of the Milky Way from 5 to 25 kpc}},
  journal = {Astrophys. J.},
  volume  = {871},
  pages   = {120},
  year    = {2019},
  eprint  = {1810.09466},
  archivePrefix = {arXiv}
}

@article{GarciaBellidoClesse2018,
  author        = {Garc{\'i}a-Bellido, Juan and Clesse, S{\'e}bastien},
  title         = {{Constraints from microlensing experiments on clustered primordial black holes}},
  journal       = {Phys. Dark Univ.},
  volume        = {19},
  pages         = {144--148},
  year          = {2018},
  doi           = {10.1016/j.dark.2018.01.001},
  eprint        = {1710.04694},
  archivePrefix = {arXiv}
}

@article{Nibauer_2026,
  author        = {Nibauer, Jacob and Bonaca, Ana and Price-Whelan, Adrian M. and Spergel, David N. and Greene, Jenny E.},
  title         = {{Measurement of Substructure from the Kinematics of the GD-1 Stellar Stream}},
  journal       = {Astrophys. J.},
  year          = {2026},
  volume        = {1004},
  pages         = {62},
  doi           = {10.3847/1538-4357/ae6776},
  eprint        = {2510.02247},
  archivePrefix = {arXiv}
}

\clearpage
\onecolumngrid
\begin{center}
\textbf{\large Supplemental Material}
\end{center}
\vspace{0.5em}
\twocolumngrid

\renewcommand{\thesection}{Supp.\Roman{section}}
\renewcommand{\theequation}{S\arabic{equation}}
\renewcommand{\thefigure}{S\arabic{figure}}
\renewcommand{\thetable}{S\arabic{table}}
\setcounter{equation}{0}
\setcounter{figure}{0}
\setcounter{table}{0}
\setcounter{section}{0}
\setcounter{secnumdepth}{2}

This Supplemental Material documents the halo, kinematic, source-population and rate/likelihood modeling behind the Letter, compares it with the published OGLE analyses, and describes the clustered-PBH calculation.

\section{Published OGLE analyses}
\label{sec:mroz_ogle}

\paragraph{20-year LMC survey.}
\citet{Mroz2024Nature} searched $\simeq80$ million LMC light curves from OGLE-III and OGLE-IV (2001--2020) and selected 13 events with objective criteria (three more were found by eye and are not in the statistical sample) \citep{Mroz2024ApJS}. Their strict and relaxed limits differ by at most $\Delta\log f\simeq0.59$, at $\log(M/M_\odot)\simeq-0.6$, and nearly coincide below $\sim10^{-4}\,M_\odot$ and above $\sim10\,M_\odot$, where stellar lenses cannot produce PBH-like timescales. Their likelihood peaks at a small nonzero abundance, $(\log M,\log f)\simeq(-1.5,-2.55)$ for the Han--Gould disk, with $\log f=-4$ disfavored by $\Delta\chi^2\simeq3.7$, driven mainly by the shortest event, OGLE-LMC-13 ($t_E\simeq7$~d). 

\paragraph{High-cadence survey.}
\citet{Mroz2024ApJL} monitored $\simeq35$ million sources in five LMC and four SMC fields (Table~\ref{tab:hc_fields}) at 16--20~min median cadence between October 2022 and May 2024. They report one long-timescale candidate, OGLE-LMC-20 ($t_E\simeq90^{+13}_{-10}$~d), and no short-timescale events. Their efficiencies $\epsilon(t_E,\rho)$ include finite-source effects and the survey cuts. They adopt the contracted MW halo, a Hernquist LMC halo of $1.49\times10^{11}\,M_\odot$, no SMC halo, and $\mathcal L=e^{-N_{\rm exp}}$, treating the one detection as stellar. PBHs from $1.4\times10^{-8}$ to $0.013\,M_\odot$ are limited to $\le1\%$ of the DM, with the minimum of the curve tabulated at $2.00\times10^{-6}\,M_\odot$.

\paragraph{Long-timescale SMC survey.}
\citet{Mroz2025ApJS} built $\sim20$-yr light curves of $\simeq7.4$ million SMC stars, found five events consistent with SMC self-lensing and MW-disk lenses, and published strict and relaxed limits.  We do not reanalyze this sample.

\paragraph{Bulge hint.}
\citet{Perkins2025} fit OGLE-IV bulge events with a hierarchical mixture that includes a DM-like lens component with a binned mass function, and find $\ln\mathcal B=20.23$ in favor of a component with masses $0.08$--$2.9\,M_\odot$. The preference is driven mainly by that component's velocity distribution. The authors caution that free-floating planets, brown dwarfs, binary lenses or $\pi_E$ selection could produce it. We treat it as an unconfirmed anomaly, not as support for our benchmark.

\paragraph{Mass function.}
All curves here and in the published analyses assume a monochromatic PBH mass function. An extended mass function redistributes events in timescale (and, toward the bulge, parallax); whether the limit at fixed total abundance weakens or strengthens depends on that distribution, the survey response and the observed events, and cannot be obtained by a universal rescaling.

\section{Model assumptions compared with the published analysis}
\label{sec:compare_mroz}

\begin{table*}[t]
\centering
\caption{Model assumptions in the published OGLE analysis \citep{Mroz2024Nature} and in our benchmark and stress test. The published MW stellar-disk entry is the Han--Gould alternative, whose relaxed limit is our reference curve. The event-origin and likelihood rows describe the 20-year LMC analysis. Enclosed LMC halo masses exclude stars and gas.}
\label{tab:assumptions}
\small
\setlength{\tabcolsep}{3pt}
\renewcommand{\arraystretch}{1.15}
\begin{ruledtabular}
\begin{tabular}{@{}
>{\raggedright\arraybackslash}p{0.22\textwidth}
>{\raggedright\arraybackslash}p{0.24\textwidth}
>{\raggedright\arraybackslash}p{0.24\textwidth}
>{\raggedright\arraybackslash}p{0.24\textwidth}
@{}}
Ingredient & Mr\'oz et al. & Benchmark & Stress test \\
\hline
MW halo density & Spherical, contracted \citep{Cautun2020} & Oblate double power law, Eq.~\eqref{eq:mwhalo} & Same as benchmark \\
MW density axis ratio $q_h$ & $1$ & $0.72$ & $0.60$ \\
$\rho_\odot$ [GeV\,cm$^{-3}$] & $\simeq0.33$ & $0.30$ & $0.20$ \\
LMC-induced disequilibrium & Not modeled & Evaluated (\ref{sec:kappa_sightline}); $\kappa_{\rm w}=1$ in rates & Same as benchmark \\
MW halo kinematics & Maxwellian, $\sigma_{\rm 1D}=V_0(R)/\sqrt{2}$ & Gaussian; non-rotating axisymmetric Jeans moments & Same as benchmark \\
MW stellar disks & Han--Gould thin and thick & Ibata et al.\ thin and thick; $\Sigma_\odot=31.55\,M_\odot\,{\rm pc}^{-2}$ & Same as benchmark \\
MW disk streaming & Cautun rotation curve & $V_c$ of the adopted total potential & Same as benchmark \\
LMC halo profile & Hernquist & Burkert & Cored isothermal \\
LMC halo parameters & $M_h=1.49\times10^{11}\,M_\odot$; $a=17.1\,{\rm kpc}$ & $\rho_b=0.031\,M_\odot\,{\rm pc}^{-3}$; $r_b=4.7\,{\rm kpc}$ & $\rho_{\rm iso}=0.057\,M_\odot\,{\rm pc}^{-3}$; $r_{\rm iso}=1.1\,{\rm kpc}$ \\
LMC halo outer cutoff & None & None & $r_t=15\,{\rm kpc}$ \\
LMC halo mass within $13.2\,{\rm kpc}$ [$10^{10}M_\odot$] & $2.83$ & $2.43$ & $1.00$ \\
LMC stellar mass [$10^9M_\odot$] & $2.7$ & $2.1$ & $2.1$ \\
Stellar IMF & Kroupa & Kroupa form, shallower slopes & Same as benchmark \\
Event origin & Strict: no PBH events; relaxed: stellar + PBH & Stellar + PBH (relaxed) & Same as benchmark \\
Likelihood & Extended Poisson; $t_E$-posterior average (relaxed) & Extended Poisson; $t_E$-posterior average & Same as benchmark \\
PBH mass function & $\delta$-function & $\delta$-function & $\delta$-function \\
\end{tabular}
\end{ruledtabular}
\end{table*}

The published analysis \citep{Mroz2024Nature} uses (i) the contracted MW halo of \citet{Cautun2020}, with Maxwellian halo velocities of dispersion $V_0(R)/\sqrt2$; (ii) an untruncated Hernquist LMC halo tied to the Orphan-stream mass \citep{Erkal2019}; and (iii) MW and LMC stellar disks and an LMC bar as stellar lenses. It also compares alternative disk models and publishes relaxed limits. Table~\ref{tab:assumptions} lists our changes. The MW and LMC halos both contribute to the PBH signal; their relative detected yields depend on mass, geometry, velocities and efficiency, not on the DM column alone.

\section{LMC-induced disequilibrium along the sightline}
\label{sec:kappa_sightline}

\paragraph{Model.}
We use the public expansion of the live MW--LMC simulation of \citet{Lilleengen2023}, in which the MW halo, MW disk and LMC halo are each represented by basis-function expansions evolved self-consistently with EXP \citep{Petersen2022EXP}; the MW-halo expansion is truncated at $l_{\max}=6$. The MW halo starts as a spherical NFW profile and the LMC as a Hernquist sphere of $1.25\times10^{11}\,M_\odot$. We evaluate the present-day ($t=0$) density in the frame of the MW-halo expansion, with the Sun $8.27$~kpc from the disk center, the solar radius of the simulation, which we retain here because only density ratios internal to the simulation are used (the rate calculations elsewhere adopt $R_\odot=8.178$~kpc), along $(\ell,b)=(280.47^\circ,-32.89^\circ)$ out to $D=49.6$~kpc. The simulated LMC lies at the same heliocentric distance and direction to within $\simeq1$~kpc. The reference $\rho^{\rm sph}_{\rm DM}$ is the $l=0$ term of the same expansion; as a second reference we use the monopole at $t=-2.4$~Gyr, before pericenter, at the same Galactocentric radius.

\begin{figure}[ht]
\centering
\includegraphics[width=\columnwidth]{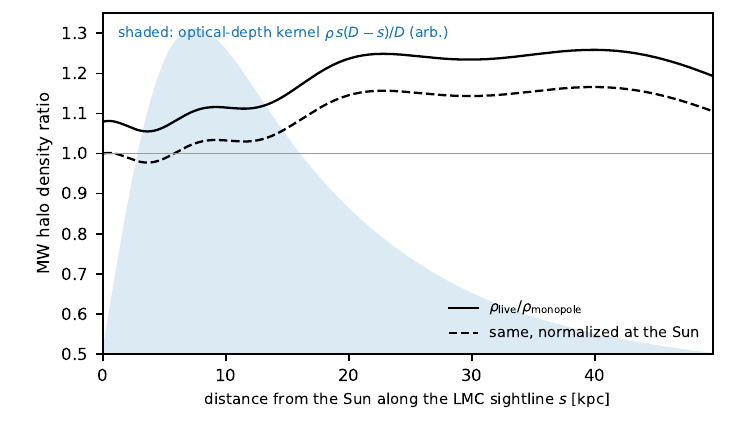}
\caption{Ratio of the present-day MW-halo density to its monopole along the sightline to the LMC in the model of \citet{Lilleengen2023} (solid), and the same ratio normalized at the Sun (dashed). The shaded curve is the optical-depth kernel $\rho\,s(D-s)/D$ in arbitrary units.}
\label{fig:kappa}
\end{figure}

\paragraph{Results.}
Figure~\ref{fig:kappa} shows $\kappa_{\rm w}(s)$ [Eq.~\eqref{eq:kappaw}]. It exceeds unity everywhere: $1.08$ at the Sun, a shallow minimum of $1.06$ near $s\simeq5$~kpc, a maximum of $\simeq1.26$ near $40$~kpc and $\simeq1.20$ at the LMC. Renormalized at the Sun it dips to $\simeq0.97$ over $s\simeq3$--$8$~kpc, where the lensing kernel is small, and exceeds unity beyond. We define
\begin{equation}
\kappa_{\rm eff}=\frac{\int_0^D ds\,\rho_{\rm DM}(s)\,W(s)}{\int_0^D ds\,\rho^{\rm sph}_{\rm DM}(s)\,W(s)},
\label{eq:kappa_eff}
\end{equation}
with $W=s(D-s)/D$ for the optical depth and $W=[s(D-s)/D]^{1/2}$ as a proxy for the event rate. Because $\rho_\odot$ is fixed by local measurements, the relevant quantity is $\kappa_{\rm eff}/\kappa_{\rm w}(0)$, with $\kappa_{\rm w}(0)=1.08$ (present monopole) or $1.03$ (pre-infall); Table~\ref{tab:kappa} lists both. Keeping only the monopole and dipole ($l\le1$) or only the monopole and quadrupole ($l\in\{0,2\}$) gives $\kappa_{\rm eff}=1.13$ and $1.02$: the dipole, which places more halo mass on the LMC side, dominates.

\begin{table}[h]
\centering
\caption{Kernel-weighted MW-halo density ratio toward the LMC in the model of \citet{Lilleengen2023}. ``Normalized'' divides by the ratio at the Sun, holding $\rho_\odot$ fixed.}
\label{tab:kappa}
\begin{ruledtabular}
\begin{tabular}{lcc}
Reference & $\kappa_{\rm eff}$ & Normalized \\
\hline
Present monopole, $\tau$ kernel & 1.147 & 1.063 \\
Present monopole, rate kernel & 1.134 & 1.050 \\
Pre-infall ($t=-2.4$~Gyr), $\tau$ kernel & 1.159 & 1.122 \\
\end{tabular}
\end{ruledtabular}
\end{table}

\paragraph{Where the optical depth comes from.}
With the kernel $\rho\,s(D-s)/D$, the fractions of the MW-halo optical depth arising beyond Galactocentric radii of $20$, $30$ and $40$~kpc are $0.25$, $0.07$ and $0.01$. The observed stellar wake and collective response lie at $30$--$100$~kpc \citep{Conroy2021,Fushimi2024}, so even an order-unity density contrast confined there changes the MW-halo optical depth by $\lesssim7\%$. Suppressing the column by tens of percent would require a comparable deficit within $\sim20$~kpc, which neither this model nor the simulations of \citet{GaravitoCamargo2019,GaravitoCamargo2021} show.

\paragraph{Limitations.}
This is a single simulation, with a spherical, uncontracted initial MW halo and an LMC lighter than the reference Hernquist halo. A more massive LMC produces a larger dipole, so the enhancement is more likely to grow than to change sign. We use only density ratios from the simulation, not its normalization. Including the enhancement would tighten our limits by $\lesssim6\%$ where the MW halo dominates the PBH yield.

\section{Published critiques and reproducibility}
\label{sec:hawkins_gb}

\paragraph{Critique and reply.}
\citet{Hawkins2025} question (i) the outer-halo normalization of the OGLE analyses, an equilibrium contracted halo extrapolated beyond the rotation-curve data, and (ii) the detection efficiencies, inferred from the deficit of observed events relative to expected self-lensing. Treating the 13 pipeline events plus three by-eye candidates as PBHs with a thermal-history mass function, they argue that $f_{\rm PBH}\sim0.8$--$1$ can be accommodated. \citet{Mroz2025Reply} respond that the efficiencies are validated by the observed $u_0$ and $t_0$ distributions and by agreement with independent bulge optical-depth measurements, that the stellar prediction (12.7--20.4 events for the 13 observed) is computed for the actual OGLE footprint, and that alternative Gaia DR3 halo models change the halo optical depth toward the LMC by only a few percent. They also cite 6D halo-tracer measurements implying $V_c\simeq180\,{\rm km\,s^{-1}}$ at the LMC distance \citep{Medina2025}.

\paragraph{Relation to this work.}
We retain the published efficiencies unchanged and do not adopt the critique's event-by-event reasoning: our likelihood, which keeps the measured timescale distribution, does not support relaxation of that magnitude. Our stress test reproduces the low outer circular speed of the models discussed by \citet{Hawkins2025}, at $50$~kpc, beyond the $\simeq27$~kpc reach of the \citet{Ou2024} data. It is in tension with \citet{Medina2025} and is presented only as an extreme case. Our benchmark agrees with the outermost rotation-curve measurement of \citet{Ou2024}.

\paragraph{Source luminosity functions.}
\label{sec:lf_inputs}
The field-specific source luminosity functions of the original analyses were not all available to us. For the high-cadence calculation, P.~Mr\'oz supplied the $I$-band luminosity functions of fields LMC510 and SMC726 (private communication), which we apply to all five LMC and all four SMC fields, respectively, including the clustered-PBH calculation. We use the supplied field- and phase-specific efficiencies for sources with $I<22$ and the corresponding foreground-corrected source counts. Reproducing the published reference limits with these inputs is a consistency check, not a proof of equivalence; the field-to-field systematic, through source weighting and finite-source sensitivity, is not quantified.

\section{The dark-matter halo of the LMC}
\label{sec:lmc_dm}

\paragraph{Constraints.}
The LMC's mass is probed by the MW-halo wake and collective response \citep{GaravitoCamargo2019,GaravitoCamargo2021,Conroy2021,Fushimi2024}, the reflex motion of the MW disk \citep{PetersenPenarrubia2021}, the Orphan-stream perturbation \citep{Erkal2019}, and globular-cluster kinematics, which give $M(<13.2\,{\rm kpc})=2.66^{+0.42}_{-0.36}\times10^{10}\,M_\odot$ \citep{Watkins_2024}. These favor a massive LMC halo but do not fix its inner profile.

\paragraph{Profiles.}
The published analyses use an untruncated Hernquist halo,
\begin{equation}
\rho_{\rm LMC}^{\rm H}(r)=\frac{M_{\rm LMC}}{2\pi}\frac{a_{\rm LMC}}{r(r+a_{\rm LMC})^3},
\label{eq:hernquist}
\end{equation}
with $M_{\rm LMC}=1.49\times10^{11}\,M_\odot$ and $a_{\rm LMC}=17.1$~kpc. Our benchmark adopts the Burkert profile
\begin{equation}
\rho_{\rm LMC}^{\rm B}(r)=\frac{\rho_b\,r_b^3}{(r+r_b)(r^2+r_b^2)},
\label{eq:burkert}
\end{equation}
with the best-fit inner-rotation-curve parameters of \citet{Regis_2021}, $\rho_b=0.031\,M_\odot\,{\rm pc}^{-3}$ and $r_b=4.7$~kpc, and no cutoff within the lensing domain. It gives $M_h(<13.2\,{\rm kpc})=2.43\times10^{10}\,M_\odot$, which together with our stellar mass is consistent with \citet{Watkins_2024}. The stress test adopts the cored isothermal fit of the same work,
\begin{equation}
\rho_{\rm LMC}^{\rm iso}(r)=\frac{\rho_{\rm iso}}{1+(r/r_{\rm iso})^2}\,\Theta(r_t-r),
\label{eq:lmc-isothermal}
\end{equation}
with $\rho_{\rm iso}=0.057\,M_\odot\,{\rm pc}^{-3}$, $r_{\rm iso}=1.1$~kpc and $r_t=15$~kpc, a truncation with precedent in LMC microlensing models \citep{vanderMarel2002,Mancini2004}. Fitted within $\simeq2.7$~kpc, these parameters give a total halo mass of $1.16\times10^{10}\,M_\odot$ and $M_h(<13.2\,{\rm kpc})=1.00\times10^{10}\,M_\odot$, requiring $\simeq1.7\times10^{10}\,M_\odot$ of baryons to match \citet{Watkins_2024}, far above our stellar mass; moving the truncation outward does not help, because the measurement lies inside it. This profile is therefore an extreme test of the LMC halo normalization, not a model consistent with all mass constraints. In both cases the core lowers the DM density along the source sightlines and is one of the ingredients behind the weakening reported in the Letter.

\section{Milky Way halo flattening and stellar distributions}
\label{sec:flattening}

\paragraph{Halo flattening.}
In Eq.~\eqref{eq:mwhalo}, $q_h=q_\rho<1$ is an oblate density. \citet{Ibata2024} obtain $q_\rho=0.75\pm0.03$ for their fixed-$\beta=3$ double power law, and \citet{Bienayme2024} $q_\rho=0.781\pm0.055$ (with potential flattening $0.843\pm0.035$). Our benchmark $q_h=0.72$ sits at the $1\sigma$ lower edge of the former. \citet{EvansBowden2014} found that, for their model family, halos flatter than $q\simeq0.57$ at the solar circle cannot reproduce the local vertical kinematics and rotation curve together; our stress-test $q_h=0.60$ is close to that edge. The tilted, triaxial halo suggested by GD-1 \citep{NibauerBonaca2025} is not captured by our disk-aligned axisymmetric model. None of these studies supplies a joint confidence region for $\rho_\odot$, $q_h$ and the kinematics adopted here. 

We quantify halo flattening through the MW dark-matter column toward the LMC center, \begin{equation} \Sigma_{\rm DM}(q_h)=\int_0^{D_{\rm LMC}} \rho_{\rm MW,DM}\!\left(R(D_L),z(D_L);q_h\right)\,\mathrm{d}D_L, \label{eq:mw-flattening-column} \end{equation} where $D_L$ is heliocentric distance and $(R,z)$ are Galactocentric cylindrical coordinates along $(\ell,b)=(280.47^\circ,-32.89^\circ)$. We take $D_{\rm LMC}=49.6\,\mathrm{kpc}$ and $R_\odot=8.178\,\mathrm{kpc}$, holding $\rho_\odot=0.30\,\mathrm{GeV\,cm^{-3}}$ and the benchmark radial-profile parameters fixed. Table~\ref{tab:mw-flattening} compares the column and detected PBH yields with their spherical counterparts; $\Sigma_{\rm DM}(1)=105.04\,M_\odot\,\mathrm{pc}^{-2}$. At $q_h=0.72$, the MW column and MW PBH yield decrease by $11.8\%$ and $19.0\%$, respectively; including the unchanged LMC contribution, the total yield falls by $13.6\%$.

\begin{table}[htbp]
\centering
\caption{Effect of MW halo flattening at fixed local density and $R_\odot=8.178\,\mathrm{kpc}$. Column ratios are $\Sigma_{\rm DM}(q_h)/\Sigma_{\rm DM}(1)$; count ratios are $N_{\rm exp}^{\rm PBH}(q_h)/N_{\rm exp}^{\rm PBH}(1)$ for the indicated halo components. Expected counts refer to the smooth 20-year survey at $M_{\rm PBH}=0.45\,M_\odot$; ``total'' includes both MW and LMC halos. Other benchmark inputs, including the LMC halo, are held fixed, while the MW potential, halo Jeans moments and observer circular speed are recomputed for each $q_h$. Count ratios include field and source-depth averaging, lensing geometry, velocities and detection efficiency.}
\label{tab:mw-flattening}
\begin{tabular}{c c c c}
\hline\hline
$q_h$ & MW column & MW count & Total count \\
 & ratio & ratio & ratio \\
\hline
$0.60$ & $0.804$ & $0.696$ & $0.783$ \\
$0.72$ & $0.882$ & $0.810$ & $0.864$ \\
$0.75$ & $0.899$ & $0.834$ & $0.882$ \\
$1.00$ & $1.000$ & $1.000$ & $1.000$ \\
\hline\hline
\end{tabular}
\end{table}

\paragraph{Milky Way disks.} We adopt the thin and thick stellar disks of \citet{Ibata2024}, using the maximum-likelihood parameters of their fixed-$\beta_h=3$ model in Table~2. Each component has density \begin{equation} \rho_i(R,z)=\frac{M_i}{4\pi R_{d,i}^{2}h_{z,i}}\,e^{-R/R_{d,i}}e^{-|z|/h_{z,i}}, \label{eq:ml-mw-disks} \end{equation} with $M_{\rm thin}=3.34\times10^{10}\,M_\odot$, $R_{d,\rm thin}=2.241\,\mathrm{kpc}$ and $h_{z,\rm thin}=0.348\,\mathrm{kpc}$, and $M_{\rm thick}=8.4\times10^{9}\,M_\odot$, $R_{d,\rm thick}=1.740\,\mathrm{kpc}$ and $h_{z,\rm thick}=0.858\,\mathrm{kpc}$. We also adopt their solar Galactocentric radius, $R_\odot=8.178\,\mathrm{kpc}$, in the microlensing rate calculations. With the global disk masses fixed at the values above, the combined stellar-disk surface density at the Sun is $31.55\,M_\odot\,\mathrm{pc}^{-2}$ and the midplane density is $0.0419\,M_\odot\,\mathrm{pc}^{-3}$. For the bulge and atomic and molecular gas disks, we follow \citet{Ibata2024} in adopting the components of \citet{McMillan2017}. The total gravitational potential combines these components with the Ibata stellar disks and our adopted dark-matter halo.

In the Han--Gould alternative of \citet{Mroz2024Nature}, the lighter disks are used as lenses while disk streaming and halo dispersion are taken from the Cautun rotation curve. In our calculation $V_c$ follows from the adopted potential, so the disks also set the streaming velocities. With our halos, Han--Gould disks give $V_c(R_\odot)=190.7$ and $173.4\,{\rm km\,s^{-1}}$ for the benchmark and stress test, far below measured values; the Ibata disks raise these to $215.2$ and $200.1\,{\rm km\,s^{-1}}$. At fixed kinematics and IMF, replacing Han--Gould with Ibata disks raises the detected MW-disk count by $\simeq30\%$.

\paragraph{LMC.}
We retain the LMC stellar model of \citet{Mroz2024Nature}: an exponential disk,
\begin{equation}
\rho_d(R'_d,z'_d)=\frac{M_d}{4\pi R_d^2 h_d}\exp\!\left(-\frac{R'_d}{R_d}-\frac{|z'_d|}{h_d}\right),
\label{eq:lmc-disk-density}
\end{equation}
and a Gaussian bar,
\begin{equation}
\rho_b=\frac{M_b}{(2\pi)^{3/2}a_b b_b c_b}\exp\!\left[-\frac12\left(\frac{x_b'^2}{a_b^2}+\frac{y_b'^2}{b_b^2}+\frac{z_b'^2}{c_b^2}\right)\right],
\label{eq:lmc-bar-density}
\end{equation}
in coordinates centered on and aligned with each component, rescaled to a total stellar mass of $2.1\times10^9\,M_\odot$. The disk and bar scale lengths are $R_d=1.8$, $h_d=0.3$, and $(a_b,b_b,c_b)=(1.2,0.44,0.44)$~kpc, with inclination $25^\circ$ and position angle $132^\circ$, at $D_{\rm LMC}=49.59$~kpc. We retain the 0.2 bar-to-disk mass ratio of \cite{Mroz2024Nature}.

\paragraph{LMC self-lensing.}
The resulting self-lensing prediction, $2.86$ detected events versus $5.7$ in the reference calculation, includes the geometry, IMF and kinematics folded through the survey response, and enters the likelihood as a fixed component; no prior on the self-lensing count is imposed. It is a model prediction, not an observational lower bound \citep{Gyuk2000,vanderMarel2002,Mancini2004,CalchiNovati2009}. A defensible uncertainty range would require varying a consistent structural and kinematic model through the full survey response; in particular, changing the depth of the source and lens layers changes both density and lens--source separation, so its effect cannot be captured by an inverse-scale-height rescaling.

\section{Microlensing rate and likelihood}
\label{sec:rate}

For a lens of mass $M$ at distance $D_L$ and a source at $D_S$,
\begin{equation}
\theta_E^2=\frac{4GM}{c^2}\frac{D_S-D_L}{D_LD_S},\quad R_E=D_L\theta_E,\quad t_E=\frac{\theta_E}{\mu},
\label{einstein}
\end{equation}
with $\mu=|\boldsymbol\mu|$ the relative proper motion. The differential rate for one source is
\begin{equation}
\frac{d^4\Gamma}{dD_L\,dM\,d^2\boldsymbol{\mu}}=2\theta_E\,\mu\,D_L^2\,\frac{dn}{dM}(D_L,M)\,f(\boldsymbol{\mu}\,|\,D_L,D_S),
\label{eq:diffrate}
\end{equation}
with $dn/dM$ the lens number density per unit mass and $f$ the normalized two-dimensional distribution of relative proper motion, $\int_{\mathbb R^2}f\,d^2\boldsymbol\mu=1$. Averaging over source distance with weight
\begin{equation}
p_s(D_S)=\frac{\rho_s(D_S)\,D_S^{2}}{\int_0^\infty\rho_s(D_S')\,D_S'^{2}\,dD_S'},
\label{eq:ps}
\end{equation}
where $\rho_s$ is the LMC stellar density (no flux-limit weighting is applied), the timescale distribution is
\begin{equation}
\begin{aligned}
\frac{d\Gamma}{dt_E}&=\int dD_S\,p_s(D_S)\int_0^{D_S}dD_L\int dM\int d^2\boldsymbol\mu\\
&\quad\times\frac{d^4\Gamma}{dD_L\,dM\,d^2\boldsymbol\mu}\,\delta_{\rm D}\!\left(t_E-\frac{\theta_E(M,D_L,D_S)}{\mu}\right),
\end{aligned}
\label{rate}
\end{equation}
evaluated by Monte Carlo integration and inserted in Eq.~\eqref{eq:nexp}, with $\Delta T_j=4000$ or $7000$~d for fields observed in OGLE-IV only or in OGLE-III$+$IV.

\paragraph{Stellar mass function.}
We write $dn/dM=A(D_L)\,\xi(M)$, with $A$ fixed by $\rho(D_L)=\int A\,\xi\,M\,dM$ over $0.01\le M/M_\odot\le10$ as in \citet{Mroz2024Nature}, and
\begin{equation}
\xi(M)=\frac{1}{M_\odot}
\begin{cases}
a_0\,m^{-\alpha_0}, & 0.01\le m<0.08,\\
m^{-\alpha_1}, & 0.08\le m<0.5,\\
a_2\,m^{-\alpha_2}, & 0.5\le m<1,\\
a_2\,m^{-\alpha_3}, & 1\le m,
\end{cases}
\label{eq:stellar-imf}
\end{equation}
with $m=M/M_\odot$, $a_0=0.08^{\alpha_0-\alpha_1}$ and $a_2=0.5^{\alpha_2-\alpha_1}$ for continuity. \citet{Kroupa2001} gives $\alpha_0=0.3\pm0.7$, $\alpha_1=1.3\pm0.5$, $\alpha_2=2.3\pm0.3$ and $\alpha_3=2.3\pm0.7$, with ranges quoted at approximately 95\% ($\alpha_1$) and 99\% ($\alpha_2,\alpha_3$) confidence and no calibrated confidence level for the quoted range below $0.08\,M_\odot$. Treating these as Gaussian two-sided intervals, the one-sided 95\% lower endpoints, $\bar\alpha-1.645\Delta/z$ with $z=1.960$ or $2.576$, are $0.88$, $2.11$ and $1.85$; we take $\alpha_0=-0.40$, the lower end of its quoted range, without a confidence level. Taking all four endpoints together is far less probable than any one of them.

\paragraph{Velocities.}
Each lens population has a Gaussian velocity distribution. For a lens at $D_L$ and a source at $D_S$, with $x=D_L/D_S$, the lens velocity relative to the microlensing tube is
\begin{equation}
\boldsymbol v_{\rm rel}=\boldsymbol v_L-x\boldsymbol v_S-(1-x)\boldsymbol v_{\rm obs},
\label{eq:vrel}
\end{equation}
where all velocities are transverse and expressed in a common frame and sky basis. For independent lens and source motions and a fixed observer velocity, the relative proper motion has mean and covariance
\begin{equation}
\overline{\boldsymbol\mu}=\frac{\overline{\boldsymbol v}_L-x\overline{\boldsymbol v}_S-(1-x)\boldsymbol v_{\rm obs}}{D_L},\qquad
\mathbf C_\mu=\frac{\mathbf C_L+x^2\mathbf C_S}{D_L^2},
\label{eq:ml-mu-moments}
\end{equation}
with $\mathbf C_L$, $\mathbf C_S$ the transverse velocity covariances, and
\begin{equation}
f(\boldsymbol\mu\,|\,D_L,D_S)=\frac{\exp\!\left[-\frac12\Delta\boldsymbol\mu^{\mathsf T}\mathbf C_\mu^{-1}\Delta\boldsymbol\mu\right]}{2\pi\sqrt{\det\mathbf C_\mu}},\qquad\Delta\boldsymbol\mu=\boldsymbol\mu-\overline{\boldsymbol\mu}.
\label{eq:ml-proper-motion-pdf}
\end{equation}

\emph{LMC sources.} In each field $j$ we take the mean stellar proper motion $\overline{\boldsymbol\mu}_{\star,j}$ and covariance $\mathbf G_j$ from the orthographic proper-motion model and tabulated RMS values of \citet{Mroz2024Nature}, transformed to the orthonormal transverse basis at the field center (RMS values as component dispersions, no correlation). The physical source moments are $\overline{\boldsymbol v}_{S,j}=D_0\overline{\boldsymbol\mu}_{\star,j}$ and $\mathbf C_{S,j}=D_0^2\mathbf G_j$, with $D_0$ the LMC distance, held fixed across source distance, so that $\mathbf C_\mu=\mathbf C_L/D_L^2+(D_0/D_S)^2\mathbf G_j$. The same source distribution is used for all lens populations.

\emph{MW disks.} Following \citet{Mroz2024Nature}, the disks stream at $\overline v_\phi=V_c(R)$, with $V_c^2=R\,\partial_R\Phi_{\rm MW}|_{z=0}$ from the adopted total potential, and have isotropic, constant dispersion $30\,{\rm km\,s^{-1}}$; we do not solve the disk Jeans equations.

\emph{MW halo.} We solve the steady axisymmetric Jeans equations for the halo tracer density $\nu_{\rm h}$, assuming $\overline v_R=\overline v_z=0$ and a velocity ellipsoid aligned with cylindrical coordinates:
\begin{equation}
\sigma_{z}^2(R,z)=\frac{1}{\nu_{\rm h}(R,z)}\int_z^\infty\nu_{\rm h}(R,z')\,\frac{\partial\Phi_{\rm MW}}{\partial z'}\,dz',
\label{eq:jeans-vertical-solution}
\end{equation}
closed by $\sigma_R^2=\sigma_z^2$, which gives
\begin{equation}
\overline{v_\phi^2}=\sigma_R^2+R\frac{\partial\Phi_{\rm MW}}{\partial R}+\frac{R}{\nu_{\rm h}}\frac{\partial(\nu_{\rm h}\sigma_R^2)}{\partial R}.
\label{eq:jeans-azimuthal-moment}
\end{equation}
With zero net rotation, $\sigma_\phi^2=\overline{v_\phi^2}$, so the covariance is ${\rm diag}(\sigma_R^2,\sigma_\phi^2,\sigma_z^2)$ in cylindrical coordinates, projected onto the sky. The potential includes the flattened halo, the Ibata disks, and the bulge and gas disks of \citet{McMillan2017}.

\emph{LMC lenses.} For self-lensing, lenses and sources are independent draws from the same field-dependent distribution, $\overline{\boldsymbol v}_{L,j}=\overline{\boldsymbol v}_{S,j}$ and $\mathbf C_{L,j}=\mathbf C_{S,j}$. LMC-halo PBHs have an isotropic dispersion
\begin{equation}
\sigma_{\rm h,LMC}(r)=\sqrt{\frac{GM_{\rm h,LMC}(<r)}{2r}},
\label{eq:lmc-halo-dispersion}
\end{equation}
with $r$ the distance from the LMC center, so $\mathbf C_L=\sigma_{\rm h,LMC}^2\mathbf I_2$. Heliocentric and Galactocentric means are transformed to a common frame before Eq.~\eqref{eq:ml-mu-moments} is applied.

\paragraph{High-cadence survey.}
Finite-source effects matter, and
\begin{equation}
N_{\rm exp}=\sum_j N_{s,j}\,\Delta T\int\epsilon_j(t_E,\rho)\,\frac{d^2\Gamma_j}{dt_E\,d\rho}\,dt_E\,d\rho,
\label{eq:expected-events-high-cadence}
\end{equation}
with $\rho=\theta_\star/\theta_E$, evaluated as in \citet{Mroz2024ApJL}. Here $\Delta T=700$~d is paired with efficiencies whose injected peak times span the same window; the shorter high-cadence coverage of each cloud (435 nights for the LMC, 236 for the SMC) is already absorbed in $\epsilon_j$.

\paragraph{Posterior.}
The likelihood is Eq.~\eqref{eq:likelihood}, with the all-population rate of Eq.~\eqref{eq:ratesum}. The inner sum averages the rate over the $2\times10^5$ $t_E$ posterior samples per event of \citet{Mroz2024Nature} [their Eq.~(6)] with equal weights. We sample the posterior in $\log_{10}f_{\rm PBH}$ by Markov chain Monte Carlo, with a prior uniform in $f_{\rm PBH}\in[0,10]$ (i.e., including the Jacobian $f_{\rm PBH}\ln10$), and report the 95th percentile of $f_{\rm PBH}$.
\paragraph{One-ingredient ablations.}
We evaluate the smooth 20-year likelihood at $M_{\rm PBH}=0.45\,M_\odot$, a typical stellar-lens mass at which PBH and stellar timescales overlap, so that every halo and stellar ingredient affects the likelihood, after restoring one reference ingredient at a time in the benchmark. We define $R_j=f_{95}^{(j)}/f_{95}^{\rm bench}$, where $(j)$ denotes the benchmark with only ingredient $j$ restored; $R_j<1$ means that the adopted ingredient weakens the bound in the context of the other benchmark choices. The observed events, field exposures, detection efficiencies and event-timescale posterior samples are identical in every calculation. We use the full 13-event likelihood, not a rescaling of expected counts or optical depth. For these one-dimensional tests the 95th percentile is obtained by direct numerical integration of the likelihood in $f$, with the same uniform prior as in the main analysis, rather than by Markov chain Monte Carlo.

For the density and flattening tests we retain the benchmark's double-power-law radial profile and change only $\rho_\odot$ or $q_h$ to the reference value listed in Table~\ref{tab:assumptions}. The MW disk change replaces the Ibata disk densities with the Han--Gould thin and thick disks. These three changes each rebuild the total potential and its dependent streaming, observer velocity and halo Jeans moments, while retaining the adopted kinematic closures. The LMC halo change uses the reference Hernquist profile and its corresponding halo dispersion. The LMC stellar-mass test rescales the disk and bar together, retaining their geometry and mass ratio, the observed source counts and empirical kinematics. The IMF test restores $(\alpha_0,\alpha_1,\alpha_2,\alpha_3) =(0.3,1.3,2.3,2.3)$ in both stellar populations, at fixed stellar mass density and with the same raw-IMF prescription and $0.01$--$10\,M_\odot$ lens-mass range.

\begin{table*}[t]
\centering
\caption{One-ingredient reversions from the benchmark in the smooth 20-year analysis at $M_{\rm PBH}=0.45\,M_\odot$, using $R_\odot=8.178\,{\rm kpc}$ in every case. For each row, $f_{95}^{(j)}$ is the recomputed 95\% upper limit after restoring the indicated reference ingredient, with the remaining benchmark inputs fixed and dependent kinematics updated consistently. We define $R_j=f_{95}^{(j)}/f_{95}^{\rm bench}$: $R_j<1$ means that the reversion tightens the limit, whereas $R_j>1$ means that it weakens the limit. The final two rows restore the corresponding ingredients together.}
\label{tab:ablation}
\small
\setlength{\tabcolsep}{6pt}
\renewcommand{\arraystretch}{1.15}
\begin{tabular}{llrr}
\hline\hline
Ingredient & Benchmark $\to$ restored value & $f_{95}^{(j)}$ & $R_j$ \\
\hline
Benchmark & Unchanged & 0.0261 & 1.00 \\
Local DM density & $0.30\to0.33\,{\rm GeV\,cm^{-3}}$ & 0.0244 & 0.94 \\
MW halo axis ratio & $q_h:0.72\to1$ & 0.0230 & 0.88 \\
LMC halo & Burkert $\to$ Hernquist & 0.0239 & 0.92 \\
MW stellar disks & Ibata $\to$ Han--Gould & 0.0286 & 1.10 \\
LMC stellar mass & $(2.1\to2.7)\times10^9M_\odot$ & 0.0257 & 0.98 \\
Stellar IMF & Adopted $\to$ canonical Kroupa & 0.0178 & 0.68 \\
\hline
All halo ingredients & $\rho_\odot$, $q_h$, LMC halo & 0.0198 & 0.76 \\
All stellar ingredients & MW disks, LMC stellar mass, IMF & 0.0202 & 0.77 \\
\hline\hline
\end{tabular}
\par\smallskip
\begin{minipage}{\textwidth}
\footnotesize
The IMF reversion sets $(\alpha_0,\alpha_1,\alpha_2,\alpha_3)=(0.3,1.3,2.3,2.3)$ in both stellar populations at fixed stellar mass density. The LMC stellar-mass reversion rescales the disk and bar together at fixed geometry. These tests retain the benchmark's MW radial halo law and kinematic prescriptions; the ratios do not provide a multiplicative decomposition of the comparison with the published OGLE limit. 
\end{minipage}
\end{table*}

The retained radial halo law, velocity prescription, LMC bar and stellar population implementation mean that none of these reversions is a complete reproduction of the published OGLE model. In particular, the tabulated factors must not be multiplied to reconstruct the benchmark-to-OGLE ratio. The IMF gives the largest individual change, $R_{\rm IMF}=0.68$, whereas the adopted MW disks act in the opposite direction, $R_{\rm disk}=1.10$. The grouped halo and stellar effects, $0.76$ and $0.77$, are comparable in these tests; they do not establish dominance of the combined stellar changes.
\section{High-cadence sample: scope and comparison}
\label{sec:hc_scope}

\begin{table}[h]
\centering
\caption{Fields of the OGLE high-cadence Magellanic Cloud survey \citep{Mroz2024ApJL}: number of sources brighter than $I=22$. High-cadence coverage is per cloud: 435 nights (LMC) and 236 nights (SMC), within the same 700~d window.}
\label{tab:hc_fields}
\begin{ruledtabular}
\begin{tabular}{lc}
Field & $N_s\,(10^6)$ \\
\hline
LMC502 & 3.26 \\
LMC503 & 8.21 \\
LMC509 & 3.77 \\
LMC510 & 5.99 \\
LMC516 & 6.36 \\
\hline
LMC total & 27.59 \\
\hline
SMC719 & 2.44 \\
SMC720 & 2.00 \\
SMC725 & 0.79 \\
SMC726 & 2.07 \\
\hline
SMC total & 7.30 \\
\end{tabular}
\end{ruledtabular}
\end{table}

\paragraph{Scope.}
The LMC fields hold 79\% of the sources and the SMC fields 21\%. The published expected count sums both clouds, includes MW-halo and LMC-halo lenses toward the LMC and MW-halo lenses toward the SMC, and neglects the SMC halo; source radii are normalized to the red clump separately for each cloud \citep{Mroz2024ApJL}. We follow the same conventions.

\paragraph{Limits as ratios of expected counts.}
At planetary masses the only detected event, OGLE-LMC-20, contributes negligible PBH rate, so Eq.~\eqref{eq:likelihood} reduces to $\mathcal L\propto e^{-f_{\rm PBH}N^{\rm PBH}_{\rm exp}(1,M)}$. With a flat prior on $[0,f_{\max}]$ and $f_{\max}N^{\rm PBH}_{\rm exp}\gg1$,
\begin{equation}
f_{95}(M)\simeq\frac{-\ln0.05}{N^{\rm PBH}_{\rm exp}(1,M)}\simeq\frac{3.0}{N^{\rm PBH}_{\rm exp}(1,M)},
\label{eq:f95}
\end{equation}
so the ratio of two limits is the inverse ratio of their expected PBH counts.


\section{Clustered primordial black holes}
\label{sec:clustered-pbhs}

\begin{figure*}[t]
\centering
\includegraphics[width=0.8\textwidth]{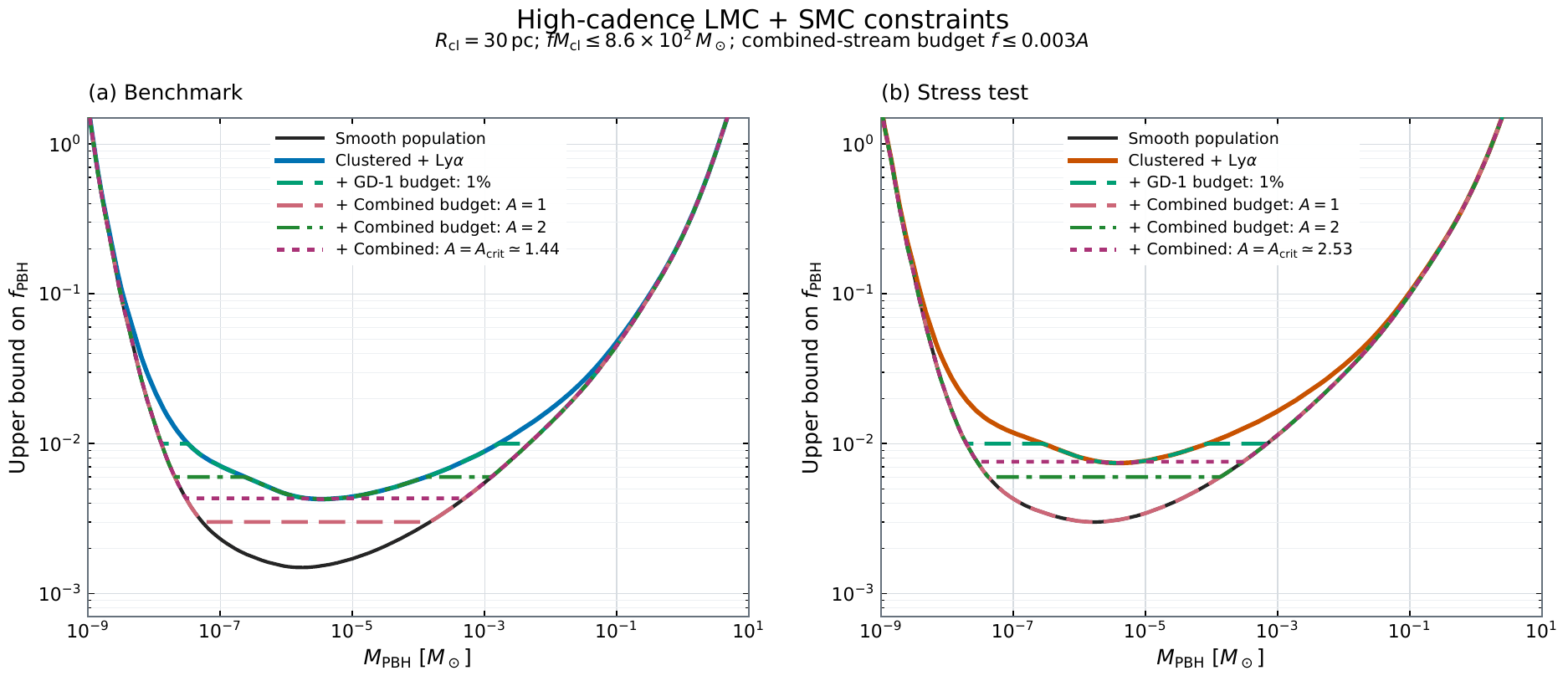}
\caption{High-cadence limits for the benchmark (a) and stress test (b), including LMC and SMC fields. Black: unclustered limits. Solid colored: clustered envelopes with the Ly$\alpha$ threshold. Other curves add the GD-1 budget $f\le0.01$ or combined-stream budgets $f\le0.003A$, with $A$ fixed along each curve; $A_{\rm crit}$ is evaluated at $2.00\times10^{-6}\,M_\odot$. Uniform clusters with $R_{\rm cl}=30$~pc. The underlying microlensing limits are 95th percentiles of the cluster-marginalized one-event posterior; the envelopes combine separate thresholds. }
\label{fig:clustered-high-cadence}
\end{figure*}

\begin{figure*}[t]
\centering
\includegraphics[width=0.8\textwidth]{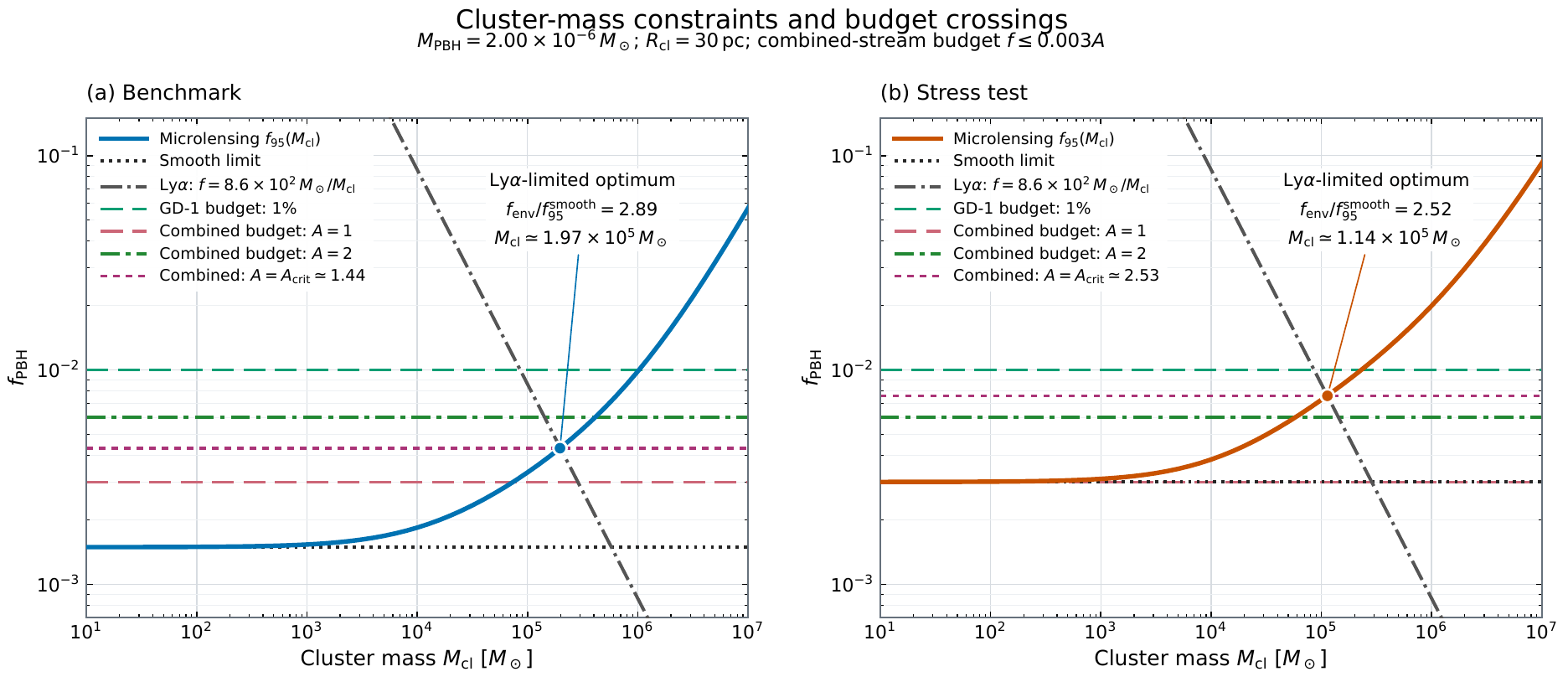}
\caption{Construction of the clustered envelopes at $M_{\rm PBH}=2.00\times10^{-6}\,M_\odot$ for the benchmark (a) and stress test (b). Rising curves: 95\% microlensing limits versus cluster mass. Descending line: Ly$\alpha$ threshold. Horizontal lines: unclustered limit and stream budgets. Markers: Ly$\alpha$-limited optima; annotations give the weakening relative to each model's unclustered limit. }
\label{fig:clustered-crossings}
\end{figure*}

Figure~\ref{fig:clustered-high-cadence} shows the resulting envelopes as functions of $M_{\rm PBH}$, and Fig.~\ref{fig:clustered-crossings} their construction at the reference mass.

\paragraph{Population.}
Spatial clustering leaves the ensemble-mean lens population unchanged but broadens the distribution of event counts across realizations \citep{GarciaBellidoClesse2018,Gorton_2022,Peta__2022}. We assume that all PBHs, of individual mass $M_{\rm PBH}$, reside in identical clusters of mass $M_{\rm cl}$ and radius $R_{\rm cl}=30$~pc, whose centers form a Poisson process with density $n_{\rm cl}=f\rho_{\rm DM}/M_{\rm cl}$, separately for the MW and LMC halos. The clusters are dilute enough that members lens individually; collective lensing by the cluster potential \citep{Toshchenko2026} is neglected.

\paragraph{Cluster yield.}
Following the projected-density convolution of \citet[Eqs.~(7)--(9)]{Peta__2022}, the fraction of a cluster projected inside field $j$ is
\begin{equation}
\eta_j(\boldsymbol x)=\frac{1}{M_{\rm cl}}\int_{\mathcal D_j}\Sigma_{\rm cl}(|\boldsymbol b-\boldsymbol b_{cj}|)\,d^2b,
\label{eq:cluster-overlap}
\end{equation}
where $\Sigma_{\rm cl}$ is the projected cluster density, $\boldsymbol b_{cj}$ the offset of the cluster from the field center and $\mathcal D_j$ the projected field aperture; this includes partial overlaps and the overlap fraction $\hat f$ of \citet[Sec.~3.2]{Gorton_2022}. For a cluster at phase-space position $\xi=(\boldsymbol x,\boldsymbol v_{\rm cl})$ in halo $h$, the expected number of detected events in field $j$ is
\begin{equation}
\lambda_{hj}(\xi)=\frac{M_{\rm cl}}{M_{\rm PBH}}\,\frac{\eta_j(\boldsymbol x)N_{s,j}\Delta T_j}{A_j(D_L)}\left\langle2u_TR_Ev_\perp\,\epsilon_j(t_E,\rho_\star)\right\rangle_{s,j|\xi},
\label{eq:cluster-yield}
\end{equation}
generalizing Eqs.~(3.14)--(3.15) of \citet{Gorton_2022}. Here $A_j(D_L)=\pi(D_L\tan\theta_j)^2$ is the aperture area at the cluster distance, with $\theta_j$ the angular radius of a circle of the field's solid angle, so that $N_{s,j}/A_j$ is the mean source surface density; $2u_TR_Ev_\perp$ with $u_T=1$ is the physical counterpart of the kernel in Eq.~\eqref{eq:diffrate}; and $\rho_\star=\theta_\star/\theta_E$. The average is over source distance [Eq.~\eqref{eq:ps}], source velocity (\ref{sec:rate}) and source luminosity, with the finite-source efficiencies of Eq.~\eqref{eq:expected-events-high-cadence}; $v_\perp$ follows from Eq.~\eqref{eq:vrel} with lens velocity $\boldsymbol v_{\rm cl}$, shared by all cluster members. Internal velocities are neglected, being small compared with the bulk transverse velocity \citep{Gorton_2022}. The total yield $\lambda_h(\xi)=\sum_j\lambda_{hj}(\xi)$ retains correlations between fields lensed by the same cluster.

\paragraph{Likelihood.}
Following the conditional Poisson counts of \citet[Eq.~(10)]{Peta__2022}, we marginalize analytically over the cluster population, whose phase-space measure at $f=1$ is
\begin{equation}
dC_h(\xi)=\frac{\rho_h(\boldsymbol x)}{M_{\rm cl}}\,p_h(\boldsymbol v_{\rm cl}\,|\,\boldsymbol x)\,d^3x\,d^3v_{\rm cl},
\label{eq:cluster-phase-space}
\end{equation}
with $p_h$ the halo velocity distribution of \ref{sec:rate}. Let $\mathcal I_{\star,j}(t_E)$ and $\mathcal I_{hj}(t_E\,|\,\xi)$ be the stellar and single-cluster detected-event intensities per unit timescale in field $j$, so that $\lambda_{hj}=\int\mathcal I_{hj}\,dt_E$. For one event observed in field $j_{\rm obs}$ at $t_{E,\rm obs}$,
\begin{equation}
\begin{aligned}
\mathcal L(f\,|\,M_{\rm PBH},M_{\rm cl})&\propto e^{-fN_{\rm eff}}\\
&\quad\times\Bigl[\mathcal I_{\star,j_{\rm obs}}(t_{E,\rm obs})\\
&\qquad{}+f\,\mathcal I_{{\rm eff},j_{\rm obs}}(t_{E,\rm obs})\Bigr],\\
N_{\rm eff}&=\sum_h\int\left[1-e^{-\lambda_h(\xi)}\right]dC_h(\xi),\\
\mathcal I_{{\rm eff},j}(t_E)&=\sum_h\int\mathcal I_{hj}(t_E\,|\,\xi)\,e^{-\lambda_h(\xi)}\,dC_h(\xi),
\end{aligned}
\label{eq:cluster-one-event}
\end{equation}
where the $f$-independent stellar factor has been absorbed into the normalization and $e^{-\lambda_h}$ accounts for the absence of further detections from the same cluster. For $\lambda_h\ll1$, $N_{\rm eff}$ and $\mathcal I_{\rm eff}$ reduce to the ensemble-mean count and intensity, and Eq.~\eqref{eq:cluster-one-event} reduces to the one-event form of Eq.~\eqref{eq:likelihood}.

\paragraph{Twenty-year sample.}
For the 20-year survey, we generalize Eq.~\eqref{eq:cluster-one-event} to the 13 detected events. We retain all partitions of the event labels, including configurations in which multiple detections originate from the same cluster, and correlations between fields intersected by that cluster.

\paragraph{Ly$\alpha$ constraint.} Using hydrodynamic simulations of the Ly$\alpha$ forest, \citet{Murgia_2019} obtained the bound $f_{\rm PBH}M_{\rm PBH}\lesssim60\,M_\odot$ at $2\sigma$ with a Gaussian prior on the reionization redshift, applicable in the Poisson regime $f_{\rm PBH}\gtrsim0.05$. \citet{De_Luca_2022} connected this constraint to initially clustered PBHs: on scales much larger than their size, independently distributed primordial clusters behave as discrete objects of mass $M_{\rm cl}$, so the bound applies to $fM_{\rm cl}$. We follow the same mapping using the newer effective-field-theory (EFT) constraints of \citet{IvanovTrifinopoulos2026}, which extend the Poisson treatment to smaller abundances, $f\gtrsim10^{-3}$, relevant here. Using the descending branch of the solid-red 95\% exclusion boundary in their published Fig.~2 gives approximately $f_{\rm PBH}M_{\rm PBH}\simeq8.6\times10^2\,M_\odot$. Identifying the discrete-object mass with $M_{\rm cl}$, we therefore adopt $fM_{\rm cl}\leq8.6\times10^2\,M_\odot$ as our approximate Ly$\alpha$ threshold. This mapping requires primordial cluster discreteness, independently distributed cluster centers, and cluster sizes small compared with the scales probed by the forest; late aggregation of an initially unclustered population does not satisfy these conditions.
\paragraph{Stellar-stream budget.} Stream perturbations independently probe dark substructure. Assuming a cold-dark-matter (CDM)-like subhalo population, \citet{Banik_2021} infer $f_{\rm sub}=(1.4^{+1.1}_{-0.7})\times10^{-3}$ within approximately $20\,{\rm kpc}$ from GD-1 and Pal~5, with a combined 95\% upper limit $f_{\rm sub}\lesssim0.003$; their GD-1-only analysis gives $f_{\rm sub}<0.01$ at 95\%. The more recent GD-1 kinematic analysis of \citet{Nibauer_2026}, modeling subhalos over $10^5$--$10^9\,M_\odot$, infers a larger fraction, $f_{\rm sub}=0.05^{+0.08}_{-0.03}$ at 68\% in their Model~I, although this fraction is normalized to the Galactic virial mass and depends jointly on subhalo abundance and compactness. These studies motivate illustrative stream allowances, but establishing an observational upper limit for our identical PBH clusters requires a dedicated calculation. In particular, the preferred cluster masses near $10^5\,M_\odot$ lie in a mass bin whose abundance is unconstrained within the prior explored by \citet{Banik_2021}. If all PBHs reside in surviving clusters that trace the halo, their contribution to the local dark-matter substructure fraction is $f_{\rm sub}^{\rm PBH}=f_{\rm PBH}$. We therefore parametrize an assumed allowance as
\begin{equation}
f_{\rm PBH}\le f_{\rm budget}(A)=Af_0,\qquad f_0=0.003,\quad A\ge1,
\label{eq:cluster-budget}
\end{equation}
where $A=1$ and $A=10/3$ reproduce the numerical combined-stream and GD-1-only allowances, respectively, and larger $A$ permits a greater clustered fraction. Here $A$ parametrizes our assumed budget; it is not inferred from the stream data.

\paragraph{Envelopes.}
For each $M_{\rm PBH}$, the Ly$\alpha$-only envelope is
\begin{equation}
F_{\rm Ly\alpha}=\max\left\{f_{95}^{\rm sm},\ \max_{M_{\rm cl}\ge M_{\rm PBH}}\min\left[f_{95}^{\rm cl}(M_{\rm cl}),\frac{(fM_{\rm cl})_{\max}}{M_{\rm cl}}\right]\right\},
\label{eq:cluster-envelope-lya}
\end{equation}
with $10^{-2}\le M_{\rm cl}/M_\odot\le10^8$, and including the stream budget,
\begin{equation}
F_{\rm Ly\alpha+streams}(A)=\max\left\{f_{95}^{\rm sm},\ \min\left[F_{\rm Ly\alpha},\,Af_0\right]\right\}.
\label{eq:cluster-envelope-streams}
\end{equation}
The outer maximum retains the unclustered population as an alternative; the stream budget applies only to the clustered branch. These envelopes combine separately imposed thresholds and are not joint 95\% posterior limits.

\paragraph{Clustered results.}
At $M_{\rm PBH}=2.00\times10^{-6}\,M_\odot$, the unclustered limits are $1.491\times10^{-3}$ and $3.004\times10^{-3}$ for the benchmark and stress test, respectively, corresponding to factors $1.19$ and $2.39$ relative to the published limit. With $(fM_{\rm cl})_{\max}=8.6\times10^2\,M_\odot$ and without imposing a stream budget, $F_{\rm Ly\alpha}=4.317\times10^{-3}$ and $7.577\times10^{-3}$, at $M_{\rm cl}\simeq1.97\times10^5$ and $1.14\times10^5\,M_\odot$: factors $2.89$ and $2.52$ beyond each model's unclustered limit, or $3.44$ and $6.03$ relative to the published limit. The stream allowance stops binding above $A_{\rm crit}=F_{\rm Ly\alpha}/f_0\simeq1.44$ and $2.53$, so $A=10/3$ does not bind at this mass. An allowance produces a plateau only where it lies between the unclustered limit and the Ly$\alpha$ envelope: at $A=1$ the benchmark envelope is $0.003$, a factor $2.01$ above its unclustered limit, while for the stress test the allowance lies slightly below the unclustered limit, which is then selected (Table~\ref{tab:cluster-budget-results}). Taking $f_{\rm budget}=0.05$ as an illustrative allowance motivated by the larger substructure fraction inferred by Nibauer et al.\ gives $A\simeq16.7$, well above both critical values, and would therefore leave both envelopes unchanged at this PBH mass.
\begin{table}[t]
\centering
\caption{Total weakening relative to the published high-cadence limit $f_{95}^{\rm ref}=1.256\times10^{-3}$ at $M_{\rm PBH}=2.00\times10^{-6}\,M_\odot$, for $R_{\rm cl}=30$~pc, $(fM_{\rm cl})_{\max}=8.6\times10^2\,M_\odot$ and illustrative stream allowance $f\le0.003A$, retaining the unclustered alternative. These ratios differ from the ratios to each model's own unclustered limit shown in Fig.~\ref{fig:clustered-crossings}. The stream allowances are not observational upper limits established for the adopted PBH clusters.}
\label{tab:cluster-budget-results}
\begin{ruledtabular}
\begin{tabular}{lcc}
Constraint & Benchmark & Stress test \\
\hline
Unclustered & $1.19$ & $2.39$ \\
Ly$\alpha$ only & $3.44$ & $6.03$ \\
Ly$\alpha$ + streams, $A=1$ & $2.39$ & $2.39$ \\
Ly$\alpha$ + streams, $A=1.5$ & $3.44$ & $3.58$ \\
Ly$\alpha$ + streams, $A=2$ & $3.44$ & $4.78$ \\
Ly$\alpha$ + streams, $A=2.5$ & $3.44$ & $5.97$ \\
Ly$\alpha$ + streams, $A=10/3$ & $3.44$ & $6.03$ \\
\end{tabular}
\end{ruledtabular}
\end{table}

\paragraph{Cluster radius and profile.}
Without a stream budget, at $M_{\rm PBH}=2.00\times10^{-6}\,M_\odot$, reducing $R_{\rm cl}$ from $30$ to $0.2$~pc weakens the limits by $3.4\%$ ($4.1\%$) for the benchmark (stress test), and increasing it to $200$~pc strengthens them by $12.5\%$ ($14.8\%$). Replacing the uniform sphere by a Plummer profile,
\begin{equation}
\rho_{\rm cl}^{\rm P}(r)=\frac{3M_{\rm cl}}{4\pi a^3}\left(1+\frac{r^2}{a^2}\right)^{-5/2},
\end{equation}
as used in PBH-cluster simulations \citep{NunoSilesGarciaBellido2025} and microlensing models \citep{Toshchenko2026}, strengthens the limits by $\simeq3\%$ in both models. The Plummer cluster has the same total mass $M_{\rm cl}$ as the uniform sphere, and its scale radius is fixed by matching the three-dimensional half-mass radius, $a=2^{-1/3}(2^{2/3}-1)^{1/2}R_{\rm cl}\simeq0.61\,R_{\rm cl}$.

\end{document}